\documentstyle[12pt]{article}

\newcommand{\sect}[1]{\setcounter{equation}{0}\section{#1}}

\def\bseq{\begin{subequation}}  % = 1a 1b
\def\eseq{\end{subequation}}
\def\bsea{\begin{subeqnarray}}  % = 1.1a 1.1b
\def\esea{\end{subeqnarray}}

\def\Bar#1{\overline{#1}}                       % big bar

\def\Tilde#1{\widetilde{#1}}                    % big tilde

\def\caja{\mathsurround=0pt}
\def\eqalign#1{\,\vcenter{\openup2\jot \caja
	\ialign{\strut \hfil$\displaystyle{##}$&$
	\displaystyle{{}##}$\hfil\crcr#1\crcr}}\,}
\def\fracm#1#2{\hbox{\large{${\frac{{#1}}{{#2}}}$}}}

\def\Dot#1{{\kern0.5pt
     {#1} \kern-5.05pt \raise5.8pt\hbox{$\textstyle.$}\kern
0.5pt}}

\newcommand{\beq}{\begin{equation}}
\newcommand{\eeq}{\end{equation}}
\newcommand{\bea}{\begin{eqnarray}}
\newcommand{\eea}{\end{eqnarray}}
\newcommand{\ena}{\end{eqnarray}}

\renewcommand{\a}{\alpha}
\renewcommand{\b}{\beta}

\renewcommand{\d}{\delta}
\newcommand{\q}{\theta}

\newcommand{\pa}{\partial}

\newcommand{\G}{\Gamma}

\newcommand{\e}{\epsilon}

\renewcommand{\l}{\lambda}
\renewcommand{\L}{\Lambda}

\newcommand{\p}{\pi}

\newcommand{\Phib}{\bar{\Phi}}

\newcommand{\chib}{\bar{\chi}}

\def\Mb{\kern 2pt\mathchoice
	    {%displaystyle
	     \vbox{\hrule width10pt height 0.4pt depth 0pt
		 \kern 1.2pt\hbox{\kern -2pt$\displaystyle M$}}}
	    {%textstyle
		 \vbox{\hrule width10pt height 0.4pt depth 0pt
		 \kern 1.2pt\hbox{\kern -2pt$\textstyle M$}}}
	    {%scriptstyle \kern 0.5pt
\vbox{\hrule width6pt height 0.4pt depth 0pt
		 \kern 1.0pt\hbox{\kern -2pt$\scriptstyle M$}}}
	    {%scriptscriptstyle \kern 0.5pt
		 \vbox{\hrule width5pt height 0.4pt depth 0pt
		 \kern 0.8pt\hbox{\kern -2pt$\scriptscriptstyle M$}}}}

\def\Sb{\kern 2pt\mathchoice
	    {%displaystyle
		 \vbox{\hrule width6pt height 0.4pt depth 0pt
		 \kern 1.2pt\hbox{\kern -2pt$\displaystyle S$}}}
	    {%textstyle
		 \vbox{\hrule width6pt height 0.4pt depth 0pt
		 \kern 1.2pt\hbox{\kern -2pt$\textstyle S$}}}
	    {%scriptstyle
		 \vbox{\hrule width3.5pt height 0.4pt depth 0pt
		 \kern 1.0pt\hbox{\kern -2pt$\scriptstyle S$}}}
	    {%scriptscriptstyle
		 \vbox{\hrule width3pt height 0.4pt depth 0pt
		 \kern 0.8pt\hbox{\kern -2pt$\scriptscriptstyle S$}}}}

\def\Rb{\kern 2pt\mathchoice
	    {%displaystyle
		 \vbox{\hrule width5.5pt height 0.4pt depth 0pt
		 \kern 1.2pt\hbox{\kern -2.5pt$\displaystyle R$}}}
	    {%textstyle
		 \vbox{\hrule width5.5pt height 0.4pt depth 0pt
		 \kern 1.2pt\hbox{\kern -2.5pt$\textstyle R$}}}
	    {%scriptstyle
		 \vbox{\hrule width3.5pt height 0.4pt depth 0pt
		 \kern 1.0pt\hbox{\kern -2.2pt$\scriptstyle R$}}}
	    {%scriptscriptstyle
		 \vbox{\hrule width3pt height 0.4pt depth 0pt
		 \kern 0.8pt\hbox{\kern -2.2pt$\scriptscriptstyle R$}}}}

  \def\pp{{\mathchoice
	  {%displaystyle
	      \kern 1pt%
	      \raise 1pt
	      \vbox{\hrule width5pt height0.4pt depth0pt
		    \kern -2pt
		    \hbox{\kern 2.3pt
			  \vrule width0.4pt height6pt depth0pt
			  }
		    \kern -2pt
		    \hrule width5pt height0.4pt depth0pt}%
		    \kern 1pt
	   }
	    {%textstyle
	      \kern 1pt%
	      \raise 1pt
	      \vbox{\hrule width4.3pt height0.4pt depth0pt
		    \kern -1.8pt
		    \hbox{\kern 1.95pt
			  \vrule width0.4pt height5.4pt depth0pt
			  }
		    \kern -1.8pt
		    \hrule width4.3pt height0.4pt depth0pt}%
		    \kern 1pt
	    }
	    {%scriptstyle
	      \kern 0.5pt%
	      \raise 1pt
	      \vbox{\hrule width4.0pt height0.3pt depth0pt
		    \kern -1.9pt  %[e]=0.15pt
		    \hbox{\kern 1.85pt
			  \vrule width0.3pt height5.7pt depth0pt
			  }
		    \kern -1.9pt
		    \hrule width4.0pt height0.3pt depth0pt}%
		    \kern 0.5pt
	    }
	    {%scriptscriptstyle
	      \kern 0.5pt%
	      \raise 1pt
	      \vbox{\hrule width3.6pt height0.3pt depth0pt
		    \kern -1.5pt
		    \hbox{\kern 1.65pt
			  \vrule width0.3pt height4.5pt depth0pt
			  }
		    \kern -1.5pt
		    \hrule width3.6pt height0.3pt depth0pt}%
		    \kern 0.5pt%}
	    }
	}}

  \def\mm{{\mathchoice
		       {%displaystyle
			     \kern 1pt
	       \raise 1pt    \vbox{\hrule width5pt height0.4pt depth0pt
				  \kern 2pt
				  \hrule width5pt height0.4pt depth0pt}
			     \kern 1pt}
		       {%textstyle
			    \kern 1pt
	       \raise 1pt \vbox{\hrule width4.3pt height0.4pt depth0pt
				  \kern 1.8pt
				  \hrule width4.3pt height0.4pt depth0pt}
			     \kern 1pt}
		       {%scriptstyle
			    \kern 0.5pt
	       \raise 1pt
			    \vbox{\hrule width4.0pt height0.3pt depth0pt
				  \kern 1.9pt
				  \hrule width4.0pt height0.3pt depth0pt}
			    \kern 1pt}
		       {%scriptscriptstyle
			   \kern 0.5pt
	     \raise 1pt  \vbox{\hrule width3.6pt height0.3pt depth0pt
				  \kern 1.5pt
				  \hrule width3.6pt height0.3pt depth0pt}
			   \kern 0.5pt}
		       }}

\def\pd{{\kern0.5pt
		   + \kern-5.05pt \raise5.8pt\hbox{$\textstyle.$}\kern 
0.5pt}}

\def\pmd{{\kern0.5pt
		  \pm \kern-5.05pt
\raise6.3pt\hbox{$\textstyle.$}\kern1.5pt}}

\def\md{{\mathchoice
   {%displaystyle
      {{\kern 1pt - \kern-6.2pt \raise5pt\hbox{$\textstyle.$}\kern
1pt}}}
    {%textstyle
      {{\kern 1pt - \kern-6.2pt \raise5pt\hbox{$\textstyle.$}\kern
1pt}}}
    {%scriptstyle
      {\kern0.5pt - \kern-5.05pt
\raise3.4pt\hbox{$\textstyle.$}\kern0.5pt}}
    {%scriptscriptstyle
      {\kern0.5pt - \kern-5.05pt
\raise3.4pt\hbox{$\textstyle.$}\kern0.5pt}}}}

\newcommand{\ad}{{\dot{\alpha}}}

\newcommand{\Del}{\nabla}
\newcommand{\Delb}{\Bar{\nabla}}
\newcommand{\Delp}{\nabla_{+}}

\newcommand{\Delpd}{\nabla_{\pd}}

\newcommand{\Delpp}{\nabla_{\pp}}
\newcommand{\Delmm}{\nabla_{\mm}}

\newcommand{\Dp}{D_{+}}
\newcommand{\Dpd}{D_{\pd}}

\newcommand{\VEV}[1]{\left\langle #1 \right\rangle}

\newcommand{\eb}{e^{\Box}}

\begin{document}

\begin{titlepage}
{\hbox to\hsize{September 2000 \hfill
{Bicocca--FT--00--13}}}
{\hbox to\hsize{${~}$ \hfill
{BRX TH-479}}}
{\hbox to\hsize{${~}$ \hfill
{IU-MSTP/41}}}
{\hbox to\hsize{${~}$ \hfill
{McGill 00-28}}}
{\hbox to\hsize{${~}$ \hfill
{UMDEPP 01-098}}}
\begin{center}
\vglue .04in
{\Large\bf Supersymmetric Gauge Anomaly \\
with General Homotopic Paths}\footnote{Supported in 
part by National Science Foundation Grants 
PHY-98-02551, PHY-9604587}
\\[.15in]
S. James Gates, Jr.\footnote{gatess@wam.umd.edu} \\
{\it Department of Physics, University of Maryland\\ 
College Park, MD 20742-4111  USA}\\
[.1in] 
Marcus T. Grisaru, \footnote{grisaru@brandeis.edu}
\footnote{On leave from Brandeis University}
Marcia E. Knutt\footnote{knutt@physics.mcgill.ca}\\
{\it Physics Department, McGill University \\
Montreal, QC Canada H3A 2T8}
\\[.1in]
Silvia Penati\footnote{Silvia.Penati@mi.infn.it}\\
{\it Dipartimento di Fisica dell'Universit\'a di
Milano-Bicocca\\ 
and INFN, Sezione di Milano, piazza delle Scienze 3, I-20126 Milano, 
Italy}\\[.1in]
Hiroshi Suzuki\footnote{hsuzuki@mito.ipc.ibaraki.ac.jp} \\
{\it Department of Mathematical Sciences, Ibaraki University, 
Mito 310-8512, Japan}
\\[.3in]

{\bf ABSTRACT}\\[.0015in]
\end{center}

We use the method of Banerjee, Banerjee and Mitra and minimal
homotopy paths to compute the consistent gauge anomaly for several
superspace models of SSYM coupled to matter. We review the derivation
of the anomaly for $N=1$ in four dimensions and then discuss the
anomaly for two-dimensional models with $(2,0)$ supersymmetry.

${~~~}$ \newline
${~~~~}$ Keywords: Gauge theories, Chiral anomaly, Supersymmetry.

\end{titlepage}

\section{Introduction}

~~~~Over the years, the construction of the superspace {\it consistent}
(chiral) anomaly for four-dimensional supersymmetric Yang-Mills theories 
has attracted the attention of many authors \cite{PIG, CLA, NIE, GUA, 
HAR, NEM, GIR, BON,  GAR, PER, KRI, MCA,MAR,OHS,GGP}.  In particular
McArthur and Osborn \cite{MCA} have discussed the construction of the 
consistent anomaly  by applying to supersymmetric theories the method 
used by Leutwyler \cite{LEU} for nonsupersymmetric theories.   In this 
method one begins with the {\em regularized} gauge variation of the 
one-loop effective action ${\mit\Gamma}$  for chiral fields in a background 
gauge field, defined in terms of the determinant of the kinetic
operator.  Due to the regularization, $(\delta {\mit\Gamma})_{reg} \neq \delta
({\mit\Gamma}_{ reg})$ and consequently the (anomalous) gauge variation does
not satisfy  the Wess-Zumino consistency condition \cite{WZC}; one must
supplement it by the variation of a local expression in order to obtain
the consistent anomaly.  Equivalently,  the covariant current must be 
supplemented by a local term, the so-called
Bardeen-Zumino current \cite{BAR}.   The construction of the latter
involves the choice of a homotopic path which connects the gauge field
to zero. Depending on  the choice of path one obtains different, 
but equally valid, forms of the consistent anomaly.

The superspace gauge field $V$ enters in the action for matter
superfields in the form $e^V$, and various authors have used a homotopy 
of the form $ e^V \rightarrow e^{yV}$ with $y$ varying between 0 and 1. 
While this  choice has some advantages, it has one serious disadvantage:
whereas the  gauge variation $\d e^V = i(\bar{\Lambda}e^V - e^V \Lambda
)$ is relatively simple, the variation of the homotopically extended
exponential $e^{yV}$ is extremely complicated. Therefore, 
in ref. \cite{GGP}, we proposed a  different homotopy $ e^V \rightarrow
1+y(e^{V}-1)$ whose gauge variation  is simple and leads to a relatively
simple form of the consistent  anomaly.  We followed the McArthur-Osborn
procedure, constructing the  four-dimensional $N=1$ consistent anomaly
as the sum of the covariant  anomaly, obtained from   
$(\delta {\mit\Gamma})_{reg}$, augmented by consistency terms.

A few years ago, Banerjee, Banerjee and Mitra \cite{BAN} proposed a
simple field-theoretic prescription which is equivalent to the Leutwyler 
method but has the advantage of constructing in one step the consistent 
anomaly.  In this prescription one starts with a regularized form of the 
effective action before computing its variation.  Consequently the 
variation automatically satisfies the consistency condition.  This
procedure was applied to the four-dimensional $N$ =1 Yang-Mills theory
by
Ohshima, Okuyama, Suzuki and Yasuta \cite{OHS} and was shown to be
equivalent to the McArthur-Osborn result.  It still requires a choice of
homotopy, and these authors made the standard choice with $e^{yV}$ so 
that the final form of the consistent anomaly is still complicated.

In this paper we apply the procedure of refs. \cite{BAN, OHS} to a
number of superspace examples, using the "minimal" homotopy of \cite{
GGP}.  To begin with, we reconsider the anomaly for four-dimensional
$N=1$  SSYM coupled to chiral superfields. We also show how the
consistent anomalies for different homotopic paths are related by gauge
transformations of a local counterterm. We then turn to some examples in
two-dimensional theories: $N=2, (2,2)$ with chiral superfields coupled
to
$(2,2)$ SSYM, and $(2,0)$ theories with chiral scalar or chiral spinor
superfields. In the first of these examples we find, as expected, no
anomaly while the anomalies in the latter two cases, when  considered
together, cancel as expected since the corresponding multiplets appear 
in the decomposition of a $(2,2)$ chiral superfield into $(2,0)$
superfields. We also show how these $(2,0)$ anomalies, when considered 
in WZ gauge, reduce to the standard gauge anomaly for fermions in two
dimensions. We have attempted to make the paper as pedagogical as
possible, at the risk of including some well-known facts. Some of these
appear in appendices.

\section{Consistent Anomaly for N=1 SSYM in Four Dimensions}

~~~~In this section we repeat, in order to make the paper
self-contained, the calculation of the anomaly as described in ref.
\cite{OHS}. The only essential difference is the use of a different
homotopy class. We consider the gauge anomaly resulting from the 
one-loop effective action~${\mit\Gamma}[V]$ of a massless chiral
superfield coupled to an external gauge superfield
\beq
S=\int d^8z\,\Phib e^V\Phi ~~~,
\label{act}\eeq
We define the effective
action by generalizing the prescription of \cite{OHS} as follows. We
introduce the one-parameter family of  gauge superfields which
smoothly interpolates between the identity and~$e^V$
\beq
   g(y,V)~~~,\qquad y\in[0,1] ~~~,
\label{subst}
\eeq
where $g(y=0,V)=1$ and~$g(y=1,V)=e^V$. We also define the corresponding
deformed superspace Yang-Mills gauge theory by the rule
\beq
e^V\to g(y,V) ~~~,
\eeq
 substituting all appearances of~$e^V$ by~$g$.  As long as
$g^\dagger=g$,
this deformation preserves supersymmetry. The action (\ref{act}) is thus
replaced for general~$y$ by
\beq
S\to S_y=\int d^8z\,\Phib g\Phi \qquad , \qquad e^{-\mit\Gamma} \equiv
\int e^{-S} \to e^{-\mit\Gamma_y}  \equiv \int [{\cal D} \Phi ] e^{-S_y} ~~~,
\eeq
and in chiral representation 
\beq
 \Del_{\a} ~\equiv~ g^{-1} D_{\a} g \quad , \quad
\Bar{\Del}_{\dot\a} \equiv \bar{D}_{\dot\a} \quad , \quad 
i \Del_{\a \, \dot\a} ~=~ \{ \Del_\a , \Bar{\Del}_{\dot\a} \} ~~~,
\label{extder}
\eeq
These deformed covariant derivatives preserve the constraints of
the usual formulation of 4D, $N$ = 1 gauge theories for all values of
$y$. We thus have a definition of 4D, $N$ = 1 supersymmetric Yang-Mills
theory that extends throughout the familiar ``cone-construction'' 
often used in discussion of homotopies.  We also define the deformed
chiral/antichiral d'Alembertians
\bea
&& \Box_+ ~=~  \Del^2 \Bar{\Del}^2 ~+~ \Bar{\Del}^2 \Del^2 ~-~
\Bar{\Del}^{\dot\a} \Del^2 \Bar{\Del}_{\dot\a} ~=~
\Box ~-~ i {\cal W}^{\a} \Del_{\a} ~-~
\frac{i}{2} (\Del^{\a} {\cal W}_{\a}) 
~~~, \nonumber \\
&& \Box_- ~=~  \Del^2 \Bar{\Del}^2 ~+~ \Bar{\Del}^2 \Del^2 ~-~
\Del^{\a} \Bar{\Del}^2 \Del_{\a} ~=~
\Box ~-~ i \Tilde{\cal W}^{\dot\a} \Bar{\Del}_{\dot\a} ~-~
\frac{i}{2} (\Bar{\Del}^{\dot\a} \Tilde{\cal W}_{\dot\a}) 
~~~, \nonumber \\
&& {\cal W}_\a ~=~ i \bar{D}^2 (g^{-1} D_\a g) \quad , \quad \Tilde{\cal 
W}_{\dot\a} ~=~ i g^{-1} D^2( g \bar{D}_{\dot\a} g^{-1}) g ~~~, \nonumber\\
&& \Box ~=~ \frac{1}{2}\Del^{\a \ad} \Del_{\a \ad}
\label{dalembert}
\ena

The effective action of the original theory~($y=1$) is obtained as
${\mit \Gamma}={\mit\Gamma}_{y=0} + \int_0^1dy\,\partial_y{\mit
\Gamma}_y$ with, from (2.4),   $\partial_y{\mit\Gamma}_y=
\VEV{\partial_yS_y}$, namely
\beq
{\mit\Gamma}[V]=\int\nolimits_0^1dy\int d^8z\,\partial_yg(z)_{ij}
\VEV{\delta S_y\over\delta g(z)_{ij}} ~~~,
\label{regact}
\eeq
where we have set~${\mit\Gamma}_{y=0}=0$. This is a formal expression
and we have to specify a regularization for the gauge current~$\VEV{
\delta S_y/\delta g}$. We regularize it ``gauge covariantly'' by
treating 
the homotopy $g$ as we would treat the usual exponential of the gauge
superfield~$e^V$,
\bea
\VEV{\delta S_y\over\delta g(z)_{ij}}
&=&\lim_{z'\to z}\VEV{\Phi(z)\Phib (z')}_{ji}
\nonumber \\
&\equiv& \lim_{z'\to z}
\left(e^{\Box_+/M^2}1/\Box_+\overline\nabla^2\nabla^2
g^{-1}\right)_{ji}\delta^{(8)}(z-z') \nonumber\\
&=& -\lim_{z'\to z}
\left( \int_{1/M^2}^{\infty} dt e^{\Box_+t}
\overline\nabla^2\nabla^2
g^{-1}\right)_{ji}\delta^{(8)}(z-z') ~~~.
\label{current}
\eea
As derived in Appendix B, the full propagator for the chiral superfield
is
\beq
\VEV{\Phi(z) \Phib (z')}_{ji} ~=~ \d_{ij} \, \Bar{\Del}^2 \frac{1}{\Box_+}
\Del^2g^{-1}\delta^{(8)}(z-z') ~~~,
\label{prop4d}
\eeq
where now the derivative operators are deformed as in (\ref{extder})
and we have regularized the propagator by 
introducing the factor $ e^{\Box_+/M^2}$,
 $M$~ being an ultraviolet cutoff.  The equations (\ref{regact}) and
(\ref{current}) are our definition of the regularized one-loop
effective action.\footnote{The gauge covariance of (\ref{current}) 
under the {\it formal\/} gauge transformation $\widehat\delta g
=i(\bar{\Lambda} g - g\Lambda)$ consid- \newline ${~~~~~~}$erably
simplifies the calculation  of the anomaly as emphasized in \cite{OHS}. 
Note that this is {\it not\/} \newline ${~~~~~~}$the gauge
transformation
on~$g$ induced by $\delta  e^V\equiv i(\bar{\Lambda} e^V-e^V\Lambda)$.}

Let us consider an arbitrary infinitesimal variation of the gauge
superfield~$\delta V$. The associated variation of the effective
action (\ref{regact}) is given by
\bea
\delta{\mit\Gamma}[V]
&=& \int_0^1 dy \int d^8z\,\delta\partial_yg(z)_{ij}
\VEV{\frac{\delta S_y}{\delta g(z)_{ij}}  }
\nonumber\\
&&\qquad+\int_0^1dy \int d^8z\int d^8z''\,
\partial_yg(z)_{ij}\delta g(z'')_{kl}
\frac{\delta}{\delta g(z'')_{kl}}
\VEV{\frac{\delta S_y} {\delta g(z)_{ij}}}.
\label{variation1}
\eea
In the first term the variation~$\delta$ commutes with the
$y$-derivative, 
$\delta\partial_y g=\partial_y\delta g$, because $\delta$~is generated
by
$\delta= \int d^8z\,\delta V (\delta/\delta V)$ which is independent
of~$y$.
Thus after integration by parts on~$y$, we have
\bea
\delta{\mit\Gamma}[V]
&=&\int d^8z \, (\delta e^{V(z)})_{ij}\VEV{\frac{\delta S}
{\delta e^{V(z)}}}_{ij}
\nonumber\\
&&\qquad+\int_0^1dy\int d^8z\int d^8z''\,
\delta g(z'')_{kl}\partial_yg(z)_{ij}
\nonumber\\
&&\qquad\qquad\qquad\qquad\quad \times\left[
\frac{\delta}{\delta g(z'')_{kl}}
\VEV{\frac{\delta S_y}{\delta g(z)_{ij}} }
-{\frac{\delta}{\delta g(z)_{ij}}}
\VEV{\frac{\delta S_y}{\delta g(z'')_{kl}}}\right]  
\nonumber\\
&&\equiv L-{1\over16\pi^2}\int_0^1dy\,X(y) ~~~.
\label{variation2}
\eea
The first term~$L$ corresponds to the covariant gauge current and is
independent of the homotopic path. The second term~$X$, which carries 
$g$ dependence, corresponds to the Bardeen-Zumino terms. Note that 
eq. (\ref{variation2}) {\it is\/} a variation of the effective
action (\ref{regact}) and thus is automatically  consistent.

\subsection{The Covariant Anomaly}

~~~~From the expression for the propagator, the covariant term can
be written as
\bea
 L &=& \int d^8z \lim_{z' \to z}\d e^{V(z)} e^{\Box_+/M^2} \frac{1}
{\Box_+} \Bar{\Del}^2 \Del^2 e^{-V(z)} \d^{(8)}(z-z') \nonumber\\
&=&  \int d^8z \lim_{z' \to z}e^{-V}\d e^V e^{\Box_+/M^2}
\frac{1}{\Box_+} \Bar{\Del}^2 \Del^2  \d^{(8)}(z-z') ~~~,
\eea
where we have brought the factor of $e^{-V} $ around to the front since 
we are dealing essentially with a trace, both in group theory labels and 
superspace (throughout the paper we neglect the ${\rm Tr}$ symbol for
the trace on the group indices reinserting it only in the final
results). 
For a gauge transformation we have
\beq
e^{-V}\d e^V ~=~ - i(  \Lambda -  e^{-V}\bar{\Lambda}e^V) ~\equiv~
- i (\Lambda - \tilde{\Lambda} ) ~~~.
\eeq
This equation allows the introduction of a holomorphic/anti-holomorphic
separation of the variation operator \cite{GGP}
\beq
e^{-V}\d e^V ~\equiv~ e^{-V}\d_R^1 e^V ~+~ e^{-V}{\bar \d}_R^1 e^V
~\to~  
e^{-V}\d_R^1 e^V  ~=~ - i\Lambda ~~,~~ e^{-V}{\bar \d}_R^1 e^V ~=~
i \tilde{\Lambda}  ~~~.
\eeq
We consider the part dependent on $\Lambda$ -- this is equivalent to the
use of the $\d_R^1$-operator to replace the $\d$-operator. The
$\tilde{\Lambda}$ part can be obtained by hermitean conjugation. 

{}From the superspace integral we pull out a factor of $d^2 \bar{\theta} 
= \Bar{\Del}^2$.  Noting that one must take the limit $z' \to z$ 
first, and using the chirality property of the various quantities 
in the integrand, we observe that $\Bar{\Del}^2$ can only act on the 
$z'$  appearing in the $\d$-function. Using $\Bar{\Del}^2 \Del^2 
\d^{(8)}(z-z') 
\stackrel{\longleftarrow}{\Bar{\Del}^2}=\Bar{\Del}^2 \Del^2 
\Bar{\Del}^2 \d^{(8)}(z-z')= \Box_+ \Bar{\Del}^2\d^{(8)}(z-z')$ we obtain
\beq
L= -i\int d^6 z \lim_{z' \to z}\Lambda e^{\Box_+/M^2}
\Bar{\Del}^2  \d^{(8)}(z-z')  ~~~.
\eeq
We write $ \d^{(8)}(z-z') = \frac{M^4}{(2 \pi)^4} \int d^4k e^{iMk(x-x')}
\d^{(4)}(\theta - \theta')$.  The derivatives act on the $x$-variable so
that the $\exp(-iMkx')$ factor can be written in front. We then pull the
$\exp(ikx)$ factor
through the derivatives picking up factors of $Mk$ along the way from the 
space-time derivatives $\partial_a$
\cite{FUJ,KON,HAY} and subsequently take the limit $x' \to x$.  We
obtain\footnote{There is a slight subtlety here. Because the  derivative 
$\partial_a$ appears also in the covariant spinor derivative $\Del_\a =
\partial_\a +\frac{1}{2} \bar{\theta}^\ad \partial_{\a \ad}$ one also generates
exponentials of $-\frac{i}{2M}W^\a \bar{\theta}^\ad k_{\a \ad}$. One can argue,
or show explicitly, that these terms do not contribute \cite{OHS}.}
\bea
L&=&-i \frac{M^4}{(2 \pi)^4} \int d^4x d^2 \theta \Lambda \\
&&\lim_{\theta' \to \theta} \int d^4 k e^{-k^2 +i k^a \Del_a/M 
+\Box/M^2 -i W^\alpha \Del_\alpha/M^2 -i(\Del^{\alpha} W_{\alpha})
/(2M^2)} \Bar{\Del}^2 \d^{(4)} (\theta - \theta') 
\nonumber
\eea
($k^2 = \frac12 k^{\a \ad}k_{\a \ad}$). 
In the limit one obtains a zero result except from terms, in the
expansion of the exponential, that can produce a factor of $\Del^2$ 
which together with the $\Bar{\Del}^2$ remove the $\d^{(4)}(\theta -
\theta')$ 
factor, i.e. from the second order term $1/2! (W^\a \Del_\a)^2$.  These 
terms also produce a factor of $1/M^4$ which cancels the $M^4$ factor. 
One can now  take the limit $\theta' \rightarrow \theta$, remove the regulator,
$M \to \infty$, perform the
$k$-integration of the remaining $e^{-k^2}$ factor, and obtain the final
form of the covariant anomaly
\beq
L= -\frac{i}{8 \pi^2} \int d^6 z ~{\rm Tr} \Big[ \, \Lambda W^\alpha
W_\alpha \, \Big] ~+~ {\rm {h.\,c.}} ~~~.
\label{covan}
\eeq

\subsection{The Consistency Terms}

~~~~We consider
\bea
&& \int d^4z''\d g(z'')_{kl}\frac{\delta}{\delta g(z'')_{kl}} 
\VEV{\frac{\delta 
S_y}{\delta g(z)} } \nonumber \\
&&~~~~ = -\d  \lim_{z'\to 
z} \left( \int_{1/M^2}^{\infty} dt e^{\Box_+t} \overline \nabla^2
\nabla^2 g^{-1}\right)_{ij}\delta^{(8)}(z-z') ~~~,
\label{4dcons}
\ena
where the covariant derivatives correspond to the deformed theory. 
As follows from eq. (\ref{extder}), in chiral representation the
dependence on $y$ in (\ref{4dcons})
only comes from $\Box_+$ and $\Del^2$. Using
\bea
\delta g^{-1} &=& - g^{-1} \delta g g^{-1} ~~~, \nonumber \\
\delta \Del_\alpha &=& \delta (g^{-1}D_\alpha g)=
[\Del_\alpha ,g^{-1} \d g] ~~~, \nonumber \\
\d \Del^2 g^{-1} &=& - g^{-1}\d g \Del^2 g^{-1} ~~~, \nonumber\\
\delta e^{\Box_+ t} &=& \int_0^t ds e^{\Box_+s} \delta \Box_+
e^{\Box_+(t-s)} ~~~,
\eea
and the fact that when acting on a chiral quantity $\Box_+ =
\Bar{\Del}^2
\Del^2$, we obtain in (\ref{variation2})
\bea
&&\int_0^1 dy \int d^8z  \, \pa_y g(z) \d \left[\, - \int_{1/M^2}^\infty
dt e^{\Box_+t} \Bar{\Del}^2 \Del^2 g^{-1} \d^{(8)}(z-z') \, \right]
{~~~~
~~~~~~}\nonumber \\
&&{~~}= \int_0^1 dy \int d^8z  \, \pa_y g(z)\left[\, \int_{1/M^2}^\infty 
dt \,e^{\Box_+t} \Bar{\Del}^2g^{-1} \d g \Del^2 g^{-1} \right. \nonumber \\
&&{~~}\left. ~~~~~~~~~~~~~~- \int_{1/M^2}^\infty  dt \int_0^t ds
e^{\Box_+s} \Bar{\Del}^2 [ \Del^2 , g^{-1}\d g] e^{\Box_+(t-s)} \Delb^2
\Del^2 g^{-1}\right]\d^{(8)}(z-z') ~~~. {~~~}\nonumber \\
&&~~~~~~~~~~ 
\label{varia}
\eea
In the last term, only the $-g^{-1}\d g \Del^2$ term from the
commutator is relevant. The other order leads to a term which,
after using the trace property of the whole expression as well
as a change of variables $s \to t-s$, cancels a corresponding term
from the expression in (\ref{variation2}) with $z$ and $z''$ interchanged.
Then one manipulates  the last term as follows:
\bea
 &&\int_{1/M^2}^\infty  dt \int_0^t ds
e^{\Box_+s} \Delb^2  g^{-1}\d g \Del^2 e^{\Box_+(t-s)} \Delb^2
\Del^2 g^{-1} {~~~~~~~~~~~~~~~~~~~~~~~~~~} \nonumber\\
&&{~~~~~~~~~~~~}=\int_{1/M^2}^\infty  dt \int_0^t ds
e^{\Box_+s} \Delb^2  g^{-1}\d g \Del^2 \Delb^2 e^{\Del^2
\Delb^2(t-s)} \Del^2 g^{-1}  \nonumber\\
&&{~~~~~~~~~~~~}= \int_{1/M^2}^\infty  dt \int_0^t ds
e^{\Box_+s} \Delb^2  g^{-1}\d g \frac{\pa}{\pa t} e^{\Box_- (t-s)}
\Del^2 g^{-1} ~~~.
\eea
Now, one integrates by parts the $\pa_t$ derivative. One obtains one
term which cancels the first term in (\ref{varia}) and one is left with
\bea
&&- \int_0^1 dy d^8z  \int_0^{1/M^2} ds \pa_y g(z) e^{\Box_+s}
\Delb^2 g^{-1} \d g e^{\Box_-(1/M^2 -s)} \Del^2 g^{-1} \d^{(8)} (z-z')
\nonumber\\
&&=-  \int_0^1 dy d^8z  \int_0^1 d \b \frac{1}{M^2} g^{-1}\pa_y
g e^{\b \Box_+/M^2} \Bar{\Del}^2 g^{-1}\d g e^{(1-\b )\Box_-/M^2}
\Del^2 \d^{(8)} (z-z')~~~. \nonumber \\
\label{expression}
\eea
This expression, and the corresponding one with $z \leftrightarrow z''$, is
manipulated in the same manner we treated the covariant anomaly. We give
some details in Appendix D (see also \cite{OHS}).
One introduces a momentum basis for the $\d^{(4)}(x-x')$ factor,
producing an  $M^4$ factor and
factors of $Mk$ in the various exponentials. One investigates then 
what survives in the limit $M^2 \to \infty$ and $\theta' \to \theta$. 
This time factors from the expansion of the exponentials proportional 
to ${\cal W}^\a \Del_\a/M^2$, $\Del^{\a} {\cal W}_{\a}/M^2$,
$\Tilde{\cal W}^\ad \Delb_\ad /M^2$ and $\Delb^{\ad} \Tilde{\cal W}_{\ad}
/M^2$ 
give the relevant contributions, cancelling the overall $M^2$ factor.
The final result takes the form
\bea
&& X(y)~=~ -16\pi^2\left(
   \delta_1\VEV{\delta_2S_g}-\delta_2\VEV{\delta_1S_g}\right)
\nonumber\\
 %  &=& 16\pi^2\int d^8z d^8z' \int_0^1d\beta\,{1\over M^2}
 %  \lim_{z'\to z}{\rm Tr} \Big( h_2 e^{\beta\Box_+/M^2}
 %  \overline\nabla^2h_1e^{(1-\beta)\Box_-/M^2}
 %  \nabla^2\delta^{(8)}(z-z') 
%\nonumber\\
%&&~~~~~~~~~~~~~~~~~~~~~~~~~~~~~~~~~~~~~~~~~~~~~
%-(1\leftrightarrow2) \, \Big)
%\nonumber\\
   &&\buildrel{M\to\infty}\over= 2i \, \int_0^1 dy\int d^8z\,
   {\rm Tr} \, h_1\left(\left[ {\cal D}^\alpha h_2,{\cal W_\alpha}
   \right] +\left[\bar{\cal D}_{\dot\alpha}h_2,
   \widetilde{\cal W}^{\dot\alpha}\right]
   +\left\{h_2,\bar{\cal D}_{\dot\alpha}
   \widetilde{\cal W}^{\dot\alpha}\right\}\right)
\nonumber\\
   &=& 2i \, \int_0^1 dy \int d^8z\, {\rm Tr} \left(
   h_1\left[\bar{\cal D}_{\dot\alpha}h_2,
   \widetilde{\cal W}^{\dot\alpha}\right]
   -h_2\left[ {\cal D}^\alpha h_1,{\cal W_\alpha}\right]\right) ~~~,
\label{xterm}
\eea
where $\delta_1g
\equiv\delta g$ and~$\delta_2g\equiv \partial_yg$, $h_1\equiv 
g^{-1} \delta g$ and~$h_2\equiv g^{-1}\partial_yg$. Moreover 
we have defined ${\cal D}_{\a} A \equiv \{ \Del_\a , A ]$ 
for any scalar or spinor object $A$ in the adjoint representation 
of the gauge group. The covariant derivatives depend explicitly on 
$y$ according to their definition (\ref{extder}).  The result in
eq. (\ref{xterm}) thus reproduces the formulae (2.25) and~(2.28) of
\cite{MCA} for general homotopic paths.  Note that our definition 
of the effective action (\ref{regact}) and (\ref{current}) is identical 
to the effective action eq.~(2.31) of \cite{MCA}, because the variations
of both formulas coincide, as eqs. (\ref{variation2}) and (\ref{xterm})
show.

The contribution $X$ of the Bardeen-Zumino current depends on the 
homotopic path $g$. An interesting property of~$X$ is \cite{OHS}
\bea
X(y)\bigr|_{\delta=\widehat\delta}
&=&16\pi^2\partial_y\int d^8z\,\widehat\delta g(z)_{ij}
\VEV{\frac{\delta S_y}{\delta g(z)_{ij}} }
\nonumber\\
&&\buildrel{M\to\infty}\over= \, \partial_y 
\int d^6z\, {\rm Tr}i\Lambda{\cal W}^\alpha
{\cal W}_\alpha+{\rm h.c.} ~~~,
\eea
which follows from the formal gauge covariance of (\ref{current}) 
under~$\widehat\delta$ as defined in the footnote below eq. (\ref{prop4d}).  
We note 
that the right hand side of this equation is (a $y$-derivative of) the
covariant anomaly with the substitution (\ref{subst}).

Different choices of the homotopic path (\ref{subst}) lead to
different forms of the consistent anomaly but two choices $g$ and~$g'$
have to be related to each other by a gauge transformation of a local
counterterm. In fact, it is easy to find the counterterm in the present
prescription. We introduce the one-parameter deformation of the path
\beq
   \gamma(y,u)~~~,\qquad u\in[0,1] ~~~,
\label{deform2}
\eeq
such that $\gamma(y,u=0)=g$ and $\gamma(y,u=1)=g'$. We then apply the
identity (we set $\delta_1\gamma=\delta \gamma$, $\delta_2\gamma=
\partial_y\gamma$ and $\delta_3\gamma=\partial_u\gamma$)
\beq
   \delta_3\Bigl(\delta_1\VEV{\delta_2S_y}
   -\delta_2\VEV{\delta_1S_y}\Bigr)
   =\delta_1\Bigl(\delta_3\VEV{\delta_2S_y}
   -\delta_2\VEV{\delta_3S_y}\Bigr)
   +\delta_2\Bigl(\delta_1\VEV{\delta_3S_y}
   -\delta_3\VEV{\delta_1S_y}\Bigr) ~~,
\eeq
to eq. (\ref{xterm}). Integrating over~$u$, we have
\bea
   &&\delta{\mit\Gamma}[V]\bigr|_{g'}
   -\delta{\mit\Gamma}[V]\bigr|_{g} 
%\\
%   &&
=\delta\biggl({-1\over16\pi^2}\biggr)
   \int_0^1du\int\nolimits_0^1dy\int d^8z \, \label{deformG}
\\
   &&\qquad\qquad\times 2i \, {\rm Tr} \Bigl[
   \gamma^{-1}\partial_u\gamma
   \left[\overline{\cal D}_{\dot\alpha}(\gamma^{-1}\partial_y\gamma),
   \widetilde{\cal W}^{\dot\alpha}\right]
   -\gamma^{-1}\partial_y\gamma
   \left[{\cal D}^\alpha(\gamma^{-1}\partial_u\gamma),
   {\cal W_\alpha}\bigr]\right] ~~~, {~~~~~~} \nonumber
\eea
where we have used the fact that the deformation (\ref{deform2}) keeps
the
endpoints $\gamma(y=0,u)=1$ and~$\gamma(y=1,u)=e^V$ fixed. The right hand
side is indeed the variation of a local term.

A possible choice of the path in (\ref{subst}) is ~\cite{MCA,MAR,OHS}
\beq
g ~\equiv~ e^{yV} ~~~.
\label{choice1}   
\eeq
The advantage of this choice is that the resulting anomaly, as given by
the last line of (\ref{xterm}) is
automatically proportional to the anomaly coefficient~$d^{abc}=
{\rm Tr} [T^a\{T^b,T^c\}]$ where $T^a$~is the representation matrix of 
the Lie algebra,  $g$  being an element of the Lie algebra (as are, then,
$h_1$, $h_2$, ${\cal W}_{\a}$ and ${\Tilde {\cal W}}_{\ad}$).  The
disadvantage of the choice (\ref{choice1}) is that the resulting
expression of the consistent anomaly involves~$\delta g$ which cannot 
be expressed in terms of $e^V$ and geometric objects but ends up to 
be a quite complicated function of~$V$.  
%We should also note that when the quantities $h_1$, $h_2$, ${\cal W}_{\a}$
%and $\Tilde{\cal W}_{\ad}$ are elements of the Lie algebra associated with
%the gauge group, the ultimate line of (\ref{xterm}) implies that $X$ is
%strictly proportional to the $d^{abc}$--coefficient.  

In ref. \cite{GGP}, use of the 4D, $N$ = 1 supersymmetric
Yang-Mills gauge theory ``minimal'' homotopy operator
\beq
g ~\equiv~ 1 ~+~ y\, (~ e^V \, - \,1 ~) ~~~,
\label{choicemin}
\eeq
was advocated.  This homotopy is minimal in the sense that it satisfies
the requisite boundary conditions {\em {and}}  and  it satisfies
Newton's second law  as well, both with respect to the exponential of the
 gauge superfield and the homotopic coordinate $y$
\beq
{\pa^2 \,g{~}\over \pa {e^V} \pa {e^V} }  ~=~ 0  ~~~,~~~
{\pa^2 \,g{~}\over \pa y^2 }  ~=~ 0  ~~~.
\label{NwTN2} \eeq
Note that  the first of these is valid at the endpoints for all relevant
homotopies.
The minimal homotopy extends its validity for all values of $y$.  Since
the 
gauge variation of the minimal homotopy is expressible in terms of
$e^V$,
\beq
\delta g ~=~ -iy\, (e^V\Lambda-\bar{\Lambda} e^V) ~~~,
\eeq
this choice yields a simple expression (see (\ref{BGJ1})-(\ref{BGJ3})
below) of the consistent gauge anomaly~ \cite{GGP}.
In the notation of ref. \cite{GGP},
with this minimal choice  the 4D, $N$ = 1 supersymmetric 
Yang-Mills Bardeen-Gross-Jackiw consistent anomaly can be expressed as
the
imaginary part  of a superaction, ${\cal A}_{\rm BGJ} = {\cal {I}{\rm
m}} [
{\Tilde {\cal  A}}{}_{\rm BGJ} \,]$ where
\beq
\eqalign{
{\Tilde {\cal A}}{}_{\rm BGJ} &=~  \fracm 1{~ 4 \pi^2 \,}  \int 
d^4 x \, d^2 \q \, d^2 {\bar \q}~ {\cal P}( \L\,; e^V) ~~~, 
{~~~~~~~} 
} \label{BGJ1}
\eeq
with \cite{GGP}
\beq
\eqalign{ 
{\cal P}( \L\,; e^V) = & \, {\rm Tr} \Big[ \, \L \, 
\Big( \,  \G^{\a} W_{\a} - \int_0^1 dy\, y \, ( \, [ \, {\cal 
W}^{\a} \, , \, \p_{\a} \, ] \,  e^V {\cal G} + \{ \, \Tilde{
\cal W}^{\dot \a} \, , \, 1 - e^V {\cal G} \, \} \Tilde
{\p}_{\dot \a}  ~)~ \Big)~ \Big]  ~, } 
\label{BGJ2} \eeq
in terms of the quantities
\beq \eqalign{ {~~~~~~}
\p_{\a} &\equiv~  e^V \, {\cal G}^2 \, \G_{\a} {~~~~~~~~~~~~
~~~~~~\,~~~} , {~\,~~~} \Tilde{\p}_{\dot \a} ~\equiv~e^V \, {
\cal G} \, \tilde{\G}_{\dot \a} \, {\cal G} ~~~~~, {~~~~~~} \cr  
{\cal W}_{\a} &\equiv~ [ {\Bar D}{}^2 ( {\cal G} D_{\a} {\cal G
}^{-1} ) \, ] ~~~~~~~~~~~~~,~~~ {\Tilde {\cal W}}{}_{\Dot \a} ~
\equiv~  {\cal G}\, [ D{}^2 ( {\cal G}^{-1} {\Bar D}{}_{\Dot \a}
{\cal G} ) \, ] {\cal G}^{-1} ~~~, \cr 
{\cal G} &\equiv~ \Big[ ~ 1 ~+~ y \, (\, e^V \, - \,1  \,) ~ \Big]^{-1}  
~~~,~~~~ W_{\a} ~=~ {\cal W}_{\a}(y = 1)
~~~.}
\label{BGJ3}\eeq 
According to the result in eq. (\ref{deformG}) any other form for
the anomaly differs from this by terms that are the gauge variation of
a local functional.

By reducing to components the expression (\ref{BGJ1}) it is easy to prove
that the bosonic component coincides with the well known result \cite{BAR}.

\section{The (2,2) Model in Two Dimensions}

~~~~This model is described by essentially the same action and
superfields as in four dimensions. Therefore, the propagator is
still given in eq. (\ref{prop4d}) and we use 
its regularization as in (\ref{current}).  
The calculation of the anomaly follows exactly the same steps,
but with one important difference: since we are working
in two dimensions the introduction of the momentum basis
\beq
\d^{(2)}(x-x') = \frac {M^2}{(2\pi)^2} \int d^2k e^{iMk(x-x')}
~~~, \eeq
only produces the overall factor of $M^2$ instead of $M^4$ as
in four dimensions. As a consequence the complete variation of
the effective action vanishes in the limit $M \to \infty$. As
expected, there is no gauge anomaly for the (left-right symmetric)
$(2,2)$ theory in two dimensions, since the theory is not chiral.

\section{BGJ Anomaly in the (2,0) SUSY Yang--Mills Theory}

~~~~In a $(2,0)$ supersymmetric theory, matter can be described by
chiral scalar
or spinor  superfields (see Appendix B for details). We compute the
gauge
anomaly for the supersymmetric Yang--Mills theory due to the presence of
chiral  matter assigned either to a scalar or to a spinor superfield.
Since the sum of the  actions for the two cases gives rise to the 
non--anomalous $(2,2)$
theory a consistency check of our calculations  will be to show that
the expressions for their gauge anomalies sum up to zero. 

The  $(2,0)$ SSYM theory is described by two quantities (see Appendix B),
a prepotential $V$ which appears in the ``plus'' covariant derivatives 
(in chiral representation) $\Delp = e^{-V}\Dp e^V$, $\Delpd = \Dpd$,
 $\Delpp = -i\{\Delp , \Delpd \}$, and an independent gauge connection 
$\Gamma_\mm$ 
which appears in $\Del_\mm = \partial_\mm -i \Gamma_\mm$.
As in the 4D case, we use a general homotopy defined by a function 
$g(y,V)$, $y \in [0,1]$ such that $g(0,V) = 1$ and $g(1,V) = e^V$.  The
deformed theory is obtained by substituting $e^V$ with $g$ everywhere. 
In particular, the deformed covariant derivatives are
\beq
\Del_+ ~=~ g^{-1} D_+ g \quad , \quad \Del_\pd ~=~ D_\pd \quad ,
\quad i\Del_\pp ~=~ \{ \, \Del_+ \,,\, \Del_\pd ~\}  \quad ,
\quad  \Del_\mm = \partial_\mm -i \Tilde{\Gamma}_\mm ~~~.
\label{extender}
\eeq
However, the supersymmetry algebra does not constrain the homotopic
extension of the  vector connection 
$\G_{\mm} \to \Tilde{\G}_{\mm}$; {\em any} independent choice 
of $\Tilde{\G}_{\mm}$ is compatible with  the SUSY algebra. 

In the formal evaluation of the anomaly given in this section we do not 
specify the particular extension of the ``minus-minus'' vector
connection, 
thus obtaining a final answer written in terms of the homotopic object
$\Tilde{\G}_\mm$.  We postpone to the next section the general
discussion 
on the possible structure of $\Tilde{\G}_\mm$.

\subsection{The Chiral Spinor Anomaly}

~~~~We begin by studying the gauge anomaly for the $(2,0)$ SSYM due to 
chiral matter described by a spinor superfield. From the minimally
coupled
action (\ref{spinoract}) we read its homotopic extension 
\beq
S ~=~ - \int d^4z ~\bar{\chi}_\md g\, \chi_- ~~~.
\label{action}
\eeq
The effective action for the original theory can be written as
\beq
{\mit\Gamma} = \int_0^1 dy \int  d^4z \pa_y g(z) \VEV{\frac{\d S}
{\d g(z)}} ~~~.
\label{effact}
\eeq

Now we compute the gauge variation of (\ref{effact})
\beq
\d {\mit\Gamma} = \int_0^1 dy \int d^4z \left[ \pa_y \d g(z) \VEV{ 
\frac{\d S}{\d g(z)}} + \pa_y g(z)  \d \VEV{ \frac{\d S}{\d g(z)}} 
\right] ~~~,
\nonumber \\
\eeq
where in the variation of the expectation value we have to vary both 
$g$ and $\Tilde{\G}_\mm$.  
Integrating by parts on $y$ one gets the integrated term
\beq
\int d^4z \d g(z) \VEV{ \frac{\d S}{\d g(z)}} \Big|_{y =1} ~~~,
\label{cov}
\eeq
which gives the covariant anomaly, and
\bea
&&\int_0^1 dy  \int d^4z d^4z''\left[ \nonumber  \pa_y g(z) \d g(z'')
\left(\frac{\d}{\d g(z'')}\VEV{ \frac{\d S}{\d g(z)}}-
\frac{\d}{\d g(z)}\VEV{ \frac{\d S}{\d g(z'')}} \right) \right.  \\
&&~~~~~~+\left. [\pa_y g(z) \d \Tilde{\G}_\mm(z'')  - 
\d g(z)\pa_y \Tilde{\G}_\mm(z'')]    
\frac{\d}{\d \Tilde{\G}_\mm(z'')}\VEV{ \frac{\d S}{\d g(z)}} \right] ~~~,
\label{cons}
\eea
which is the consistency piece.

The expectation value in the previous expressions must be thought as 
being suitably regularized. Proceeeding as in the four dimensional case, 
we write 
\bea
\VEV{\frac{\d S}{\d g(z)}} &=& - \VEV{\frac{\d}{\d g} \int d^4z \bar{
\chi}_\md g \chi_-} = \lim_{z' \to z} \VEV{\chi_-(z) \chib_\md (z')}
\nonumber \\
&=&  i \lim_{z' \to z} e^{\Box_+ /M^2}\frac{1}{\Box_+} \Delpd \Tilde{
\Del}_\mm \Delp g^{-1} \d^{(4)}(z-z') \nonumber\\
&=&
-i \lim_{z' \to z} \int_{1/M^2}^{\infty} dt  e^{\Box_+ t} \Delpd 
\Tilde{\Del}_\mm \Delp g^{-1}\d^{(4)}(z-z') ~~~.
\label{regprop}
\eea
Here the extended expression for the spinor propagator 
(\ref{spinorprop}) has been used.

\subsubsection{The Covariant Anomaly}

~~~~From eq. (\ref{cov}) and the expression in the second line of
(\ref{regprop}) the covariant anomaly is given by (we bring the $g^{-1}$
around because we have basically a trace)
\beq
\d {\mit\Gamma}_{cov} ~=~ i \int d^4z \lim_{z' \to z} (g^{-1} \d g) \,
e^{\Box_+ /M^2}\frac{1}{\Box_+} \Delpd \Tilde{\Del}_\mm  \Delp
\d^{(4)}(z-z')\Big|_{y =1} ~~~.
\eeq
Since $g^{-1}\d g |_{y=1} = i (\tilde{\L} - \L)$ where 
$\tilde{\L} \equiv e^{-V}\bar{\L} e^V$, 
we split the above expression into a sum of two pieces proportional
to $\L$ and $\tilde{\L}$ respectively, which are  the hermitian
conjugate
of one another.  Therefore, we only concentrate on the $\L$ piece.

In the $\L$--term we pull out of the superspace integration measure a
spinor 
derivative $\Delpd$ which acts on both $z$ and $z'$ due to the limit $z'
\to 
z$.  When it acts on the $z$ variable, it is acting from the left on
chiral
objects and the result is zero.  Instead, when it acts on the $z'$
variable,
it is just acting on the $\d$-function from the right, and using the
chain of identities
$\Del_{z'} ...\d^{(4)}(z-z') =  ...
\d^{(4)}(z-z')\stackrel{ \leftarrow}{\Del}_{z'} = - ...\Del_{z}
\d^{(4)}(z-z')$.
we obtain 
\bea
\d {\mit\Gamma}_{cov} &&=
-\int d^2x d\q^+ ~\lim_{z' \to z} \L  e^{\Box_+ /M^2}
\frac{1}{\Box_+}
\Delpd \Delmm \Delp \Delpd  \d^{(4)}(z-z')\nonumber\\
&&=-i \int d^2x d\q^+ ~\lim_{z' \to z} \L
e^{\Box_+ /M^2} \Delpd   \d^{(4)}(z-z') ~~~,
\eea
where we have legitimately cancelled the $1/\Box_+$.
In the limit we will get
zero unless we can pull out of the exponential a factor of $\Delp$
to act on the $\d^{(2)}(\q -\q')$.
To this end, we write again
\beq
\d^{(4)}(z-z') =\d^{(2)} (x-x') \d^{(2)} (\q - \q') =
M^2\int \frac{d^2k}{(2 \pi)^2} \,
e^{iMk(x-x')}  \d^{(2)} (\q- \q')  ~~~,
\eeq
and pull the $e^{iMkx}$ factor through the derivatives before taking  
the limit $x' \to x$.  The result of this operation is that again  the various
derivatives are shifted  by factors of $Mk$, 
and using the explicit form of $\Box_+$ the exponential becomes
\beq
\exp \left[- k^2  + ik^a\Del_a /M+\Box/M^2 -1/2M^2(\Delpd \Bar{W}
_\md -\Delp W_-) -W_-\Delp /M^2 \right] 
\eeq
($k^2 = k_\pp k_\mm$). 
Expanding in powers of $1/M$ only the last term in the exponential 
will contribute in the limit $\theta' \rightarrow \theta$,
 $M \rightarrow \infty$, when expanded to 
first order.  The only thing that survives in the exponential is just 
the $k^2$ term so the $k$ integral can be explicitly performed and
gives a factor of $\pi$.  We obtain then, for the covariant anomaly,
\beq
\d {\mit\Gamma}_{cov} ~=~ \frac{i}{4 \pi} \int d^2x d \q^+ {\rm Tr}
\left( \L W_- \right) ~~+~~{\rm h.c.} ~~~.
\label{s1}
\eeq
This result can also be obtained by a very simple supergraph
calculation of the relevant  two-point function for the one-loop
gauge field effective action due to a chiral spinor loop.

\subsubsection{The Consistency Terms}

~~~~We now focus on the consistency term (\ref{cons}) where we use 
the last line in eq. (\ref{regprop}) for the  expectation value $\VEV{
\d S/ \d g}$. In that expression the spinorial derivatives are 
homotopically extended according to eq. (\ref{extender}) and the vector
connection $\G_\mm$ is independently extended to $\Tilde{\G}_\mm$.

We will make use of the following identities
\bea
&&\frac{\d}{\d g} \Delpd \Tilde{\Del}_\mm \Delp g^{-1} =
\frac{\d}{\d g} \Delpd \Tilde{\Del}_\mm  g^{-1} \Dp
\nonumber \\
&& ~~~=- \Delpd \Tilde{\Del}_\mm g^{-1} \frac{\d g}{\d g} g^{-1}D_+ 
= - \Delpd \Tilde{\Del}_\mm g^{-1} \frac{\d g}{\d g} \Delp g^{-1} ~~~,
\eea
so that
\bea
&&\pa_y g \frac{\d}{\d g} \Delpd \Tilde{\Del}_\mm \Delp g^{-1} =
-\Delpd \Tilde{\Del}_\mm (g^{-1} \pa_y g) \Delp g^{-1} \nonumber \\
&&\d g \frac{\d}{\d g} \Delpd \Tilde{\Del}_\mm \Delp g^{-1} = 
-\Delpd \Tilde{\Del}_\mm (g^{-1} \d g )\Delp g^{-1} ~~~.
\eea
Similarly 
\bea
&&\pa_y \Tilde{\G}_\mm \frac{\d}{\d \Tilde{\G}_\mm} \Delpd 
\Tilde{\Del}_\mm \Delp g^{-1} = -i\Delpd \pa_y \Tilde{\G}_\mm 
\Delp g^{-1} \nonumber \\ 
&&\d \Tilde{\G}_\mm \frac{\d}{\d \Tilde{\G}_\mm} \Delpd 
\Tilde{\Del}_\mm \Delp g^{-1} = -i\Delpd \d \Tilde{\G}_\mm  
\Delp g^{-1} ~~~.
\eea
Also we have
\beq
\d \int_{1/M^2}^{\infty} e^{\Box_+ t} = \int_{1/M^2}^{\infty}dt
\int_0^t ds e^{\Box_+ s} \d{\Box_+} e^{\Box_+ (t- s)} ~~~,
\eeq
where $\Box_+$ is given in (\ref{spinorbox}) and $\d \Box_+$ acting 
on a chiral expression is
\beq
\d \Box_+ \rightarrow -i \d \Delpd \Tilde{\Del}_\mm \Delp = 
-\Delpd  \d \Tilde{\G}_\mm \Delp -i \Delpd \Tilde{\Del}_\mm
[\Delp ,( g^{-1} \d g) ] ~~~.
\label{var}
\eeq
Using all the previous identities it is now easy to compute the
consistent
terms in (\ref{cons}). We begin by considering the first line in 
(\ref{cons}).  Leaving aside an overall $\int_0^1 dy \int d^4 z \lim_{z'
\to z} \d^{(4)}(z-z')$  it can be written as
\bea
&& -i(g^{-1} \d g) \int_{1/M^2}^{\infty} e^{\Box_+ t} \Delpd
\Tilde{\Del}_\mm (g^{-1}\pa_y g) \Delp +i
(g^{-1} \pa_y g) \int_{1/M^2}^{\infty} e^{\Box_+ t} \Delpd
\Tilde{\Del}_\mm (g^{-1}\d g) \Delp \nonumber \\
&&-i(g^{-1}\pa_y g)  \int_{1/M^2}^{\infty}dt \int_0^t ds
e^{\Box_+ s} \Delpd \Tilde{\Del}_\mm [ \Delp , (g^{-1}\d g)]
e^{\Box_+ (t-s)} \Delpd \Tilde{\Del}_\mm \Delp (-i) \nonumber\\
&&+i(g^{-1}\d g)  \int_{1/M^2}^{\infty}dt \int_0^t ds
e^{\Box_+ s} \Delpd \Tilde{\Del}_\mm [ \Delp , (g^{-1}\pa_y g)]
e^{\Box_+ (t-s)} \Delpd \Tilde{\Del}_\mm \Delp (-i) ~~~. \nonumber \\
\label{first}
\eea
Now, it is easy to check that from the commutators such as $ [ \Delp 
, (g^{-1}\d g)] $ only the second terms survives, since in the first 
terms one can use the cyclic property of the trace to show that the 
corresponding terms  from the last two lines in (\ref{first}) are
identical and cancel. Therefore, we are left with
\beq
\eqalign{
&-i(g^{-1} \d g) \int_{1/M^2}^{\infty} e^{\Box_+ t} \Delpd
\Tilde{\Del}_\mm (g^{-1}\pa_y g) \Delp \cr
&-i(g^{-1}\d g)  \int_{1/M^2}^{\infty}dt \int_0^t ds e^{\Box_+ s}
\Delpd \Tilde{\Del}_\mm (g^{-1}\pa_y g)\Delp e^{\Box_+ (t-s)} (-i)\Delpd
\Tilde{\Del}_\mm \Delp ~-~ \d \leftrightarrow \pa_y ~~~.
}
\label{firstII} 
\eeq
The second term of this expression can be rewritten as
\bea
&&-i(g^{-1}\d g)  \int_{1/M^2}^{\infty}dt \int_0^t ds e^{\Box_+ s}
\Delpd \Tilde{\Del}_\mm  (g^{-1}\pa_y g)\Delp e^{\Box_+ (t-s)} 
(-i)\Delpd \Tilde{\Del}_\mm \Delp \nonumber \\
&& =-i (g^{-1}\d g)  \int_{1/M^2}^{\infty}dt \int_0^t ds e^{\Box_+ s}
\Delpd \Tilde{\Del}_\mm (g^{-1}\pa_y g) \Delp \frac{\pa}{\pa t}
e^{-i\Delpd \Tilde{\Del}_\mm \Delp (t-s)} \nonumber \\
&& =-i (g^{-1}\d g)  \int_{1/M^2}^{\infty}dt \int_0^t ds e^{\Box_+ s}
\Delpd \Tilde{\Del}_\mm (g^{-1}\pa_y g) \frac{\pa}{\pa t}
e^{\Box_- (t-s)} \Delp ~~~,
\eea
where $\Box_-$ has been defined in (\ref{spinorboxm}).  Integrating the
$t$-derivative by parts one gets the integrated term, and another term
where the derivative acts on the upper limit of the $s$-integral.  This 
second term cancels the first line of (\ref{firstII}).  Therefore we are 
left with
\bea
&&i(g^{-1}\d g) \int_0^{1/M^2} ds e^{\Box_+ s} \Delpd \Tilde{\Del}_\mm 
(g^{-1} \pa_y g)e^{\Box_- (1/M^2-s)} \Delp
\nonumber \\
&& -i (g^{-1}\pa_y g) \int_0^{1/M^2} ds e^{\Box_+ s} \Delpd 
\Tilde{\Del}_\mm (g^{-1} \d g) e^{\Box_- (1/M^2-s)} \Delp ~~~.
\eea
Again, besides an integration over $y$, a functional trace is
understood;
equivalently there is a $\d^{(4)}(z-z')$ which is treated as in the 
case of the covariant anomaly. Introducing a momentum basis as before 
and sending $M^2 \to \infty$ after performing the
$k$--integration we finally obtain
\beq
-\frac{i}{4\pi}\int_0^1 d y \int d^4z \,  {\rm Tr} \left[(g^{-1} \d g) 
\Tilde{\Del}_\mm (g^{-1}\pa_y g) - (g^{-1}\pa_y g) \Tilde{\Del}_\mm 
(g^{-1} \d g)  \right] ~~~.
\label{s2}
\eeq
The second term can be integrated by parts into the first one.

This is the contribution to the consistent anomaly from the first line
in eq. (\ref{first}). Now we look at the second line. Substituting the
regularized expression (\ref{regprop}) for the expectation value we have 
\bea
&&-i(g^{-1}\pa_y g(z)) \d \Tilde{\G}_\mm (z') \frac{\d}{ \d
\Tilde{\G}_\mm
(z')} \int dt \, e^{\Box_+ t} \Delpd \Tilde{\Del}_\mm \Delp ~-~ \d
\leftrightarrow \pa_y \nonumber \\ 
&&= -(g^{-1}\pa_y g(z))\int dt \, e^{\Box_+ t} \Delpd \d \Tilde{\G}_\mm
\Delp \nonumber\\
&&+i(g^{-1}\pa_y g(z))\int dt \int ds e^{\Box_+ s} \Delpd \d 
\Tilde{\G}_\mm \Delp  e^{\Box_+ (t-s)} \Delpd \Tilde{\Del}_\mm \Delp 
\nonumber \\ 
&& - \d \leftrightarrow \pa_y ~~~.
\eea
Here we have used $\d \Tilde{\Del}_\mm = -i \d \Tilde{\G}_\mm$.  Now 
one plays standard games exactly as before.   Without giving the details 
the final result from the second line in (\ref{first}) is
\beq
-\frac{1}{4 \pi}  \int_0^1 d y \int d^4z ~{\rm Tr} \left[\, (g^{-1}\pa_y
g) 
\, \d \Tilde{\G}_\mm ~-~ (g^{-1}\d g) \pa_y \Tilde{\G}_\mm \, \right]
~~~.
\label{s3s4}
\eeq
Adding the contributions from (\ref{s1}) and (\ref{s2}) the final
expression
for the (2,0) BGJ anomaly due to a chiral spinor superfield is   
\bea
{\cal A}_{\rm BGJ} &&=~ \frac{i}{4 \pi} \int d^2x d \q^+ ~{\rm Tr}
 \left( \L W_-\right) ~-~ \frac{i}{4 \pi} \int d^2x d \q^\pd ~{\rm Tr}  
\left( \bar{\L} \Bar{W}_\md \right) 
\nonumber \\
&&~~ 
-\frac{i}{2\pi}\int_0^1 d y \int d^4z ~ {\rm Tr} \left[\, (g^{-1} \d g) 
\Tilde{\Del}_\mm (g^{-1}\pa_y g) \, \right] \nonumber \\
&&~~
-\frac{1}{4 \pi}  \int_0^1 d y \int d^4z ~ {\rm Tr} \left[ \,
(g^{-1}\pa_y g) 
\, \d \Tilde{\G}_\mm ~-~ (g^{-1}\d g) \, \pa_y \Tilde{\G}_\mm \,
\right]
~~~,
\label{BGJanomaly}
\ena
where in (\ref{s2}) an integration by parts on the second term has been
performed. Here we have defined $ \Bar{W}_\md \equiv - (W_-)^{\dag}$.

\subsection{The (2,0) Chiral Scalar Anomaly}

~~~~The evaluation of the gauge anomaly due to matter in a chiral scalar
superfield follows exactly the same procedure as in the spinor case.
Therefore, in this section we only describe the main steps of the
calculation without giving details.

The supersymmetric action for a chiral scalar coupled to a gauge field 
is given in (\ref{scalaract}).  We homotopically extend $e^V \rightarrow
g$ 
as well as $\G_\mm \rightarrow \Tilde{\G}_\mm$ obtaining
\beq
S ~=~ -i\int d^2x d^2\theta \, \bar{\Phi} g \Tilde{\Del}_\mm \Phi ~~~.
\label{scalextend}
\eeq
We write the effective action as
\beq
{\mit\Gamma} = 
\int_0^1 dy \int  d^4z \left[ \pa_y g(z) \VEV{ \frac{\d S}{\d g(z)}} +
\pa_y \Tilde{\G}_\mm(z) \VEV{ \frac{\d S}{\d \Tilde{\G}_\mm(z)}} \right]
~~~. \eeq
In contradistinction to the spinor case, now 
the action (\ref{scalextend}) depends explicitly on $\Tilde{\G}_\mm$ 
so that computing the gauge variation one gets a more complicated
expression
\bea
\d {\mit\Gamma} &=& 
\int_0^1 d y \int d^4z \left[ ~\pa_y \d g(z)\VEV{ \frac{\d S}{\d g(z)}}
+ \pa_y g(z) \d \VEV{ \frac{\d S}{\d g (z)}} \right. \cr
&& 
\left. ~~~~~
+\pa_y  \d \Tilde{\G}_\mm(z) \VEV{ \frac{\d S}{\d \Tilde{\G}_\mm (z)}} 
+ \pa_y \Tilde{\G}_\mm (z) \d \VEV{ \frac{\d S}{\d \Tilde{\G}_\mm (z)}}  
\, \right]  ~~~.
\ena
As before, we integrate by parts on $y$ obtaining the integrated term
\beq
\int d^4z \left[ \, \d g(z) \VEV{ \frac{\d S}{\d g(z)}} + \d
\Tilde{\G}_\mm(z)
\VEV{ \frac{\d S}{\d \Tilde{\G}_\mm(z)}}\right]_{y =1} ~~~,
\eeq
which represents the covariant anomaly, and
\bea
&&\int dy  \int d^4z d^4z''\left[  \pa_y g(z) \d g(z'')
\left(\frac{\d}{\d g(z'')}\VEV{ \frac{\d S}{\d g(z)}}-
\frac{\d}{\d g(z)}\VEV{ \frac{\d S}{\d g(z'')}} \right) \right. 
\nonumber\\
&&~~~~~~+\left.\pa_y g(z) \d \Tilde{\G}_\mm(z'')  \left(\frac{\d}{\d 
\Tilde{\G}_\mm(z'')} \VEV{ \frac{\d S}{\d g(z)}}-
\frac{\d}{\d \Tilde{\G}_\mm(z)}\VEV{ \frac{\d S}{\d g(z'')} } \right)
\right.  
\nonumber \\
&&~~~~~~~+\left.\pa_y \Tilde{\G}_\mm(z) \d \Tilde{\G}_\mm(z'')
\left(\frac{\d}{\d \Tilde{\G}_\mm(z'')}\VEV{ \frac{\d S}{\d \Tilde{
\G}_\mm(z)}}- \frac{\d}{\d \Tilde{\G}_\mm(z)}\VEV{ \frac{\d S}{\d
\Tilde{\G}_\mm(z'')}} \right) \right.   
\nonumber \\
&&~~~~~~+\left.\pa_y \Tilde{\G}_\mm(z) \d g(z'')  \left(\frac{\d}{\d
g(z'')}
\VEV{ \frac{\d S}{\d \Tilde{\G}_\mm(z)}}- \frac{\d}{\d g(z)}\VEV{
\frac{\d
S}{\d \Tilde{\G}_\mm(z'')} } \right)\right] ~~~,
\label{second}
\eea
which is the consistency piece.

\subsubsection{The Covariant Anomaly}

~~~~We look first at the covariant term
\bea
\d {\mit \G}_{cov} &=& \left. \int d^4z \left( \d g \VEV{\frac{\d S}{\d g}} +
\d \Tilde{\G}_\mm \VEV{\frac{\d S}{\d \Tilde{\G}_\mm}} \right) 
\right|_{y=1} \nonumber \\
&=& - \lim_{z' \to z} \int d^4z \left[ i \d g \Tilde{\Del}_\mm 
\VEV{ \Phi  (z)\Phib (z')} + \d \Tilde{\G}_\mm\VEV{ \Phi(z) \Phib 
(z') } g(z') \right] \Big|_{y=1} ~~~.
\nonumber \\
\label{cov2}
\ena
In this case
the regularized expression for the scalar propagator (\ref{scalarprop})
is given by 
\bea
&& \VEV{\Phi(z) \Phib(z')} ~=~ e^{\Box_+ /M^2}
\Del_\pd \frac{1}{\Box_+}   \Delp g^{-1} \d^{(4)}(z-z') \nonumber \\
&&~~~~~~~~~=~ - \int_{\frac{1}{M^2}}^{\infty} dt \Del_\pd e^{\Box_+ t}
\Del_+ g^{-1} \d^{(4)}(z-z') ~~~.
\label{scalarpropreg}
\ena
The expression (\ref{cov2}) is evaluated at $y =1$
so the gauge variations are the usual ones, chosen to ensure the
invariance
of the classical action (see (\ref{gvariation}))
\beq
g^{-1} \d g|_{y=1} = i \tilde{\L} - i \L ~~~~,~~~~ 
\d \Tilde{\G}_\mm|_{y=1}  = \d \G_\mm =  \Del_\mm \L ~~~,
\eeq
where again we have defined $\tilde{\L} \equiv e^{-V} \bar{\L} e^V$.

Thefore, with  some obvious integration by parts, the expression for 
the covariant anomaly can be written as
\bea
\d {\mit \G}_{cov}
&=& \int d^4z {\rm Tr} \left[ \left( -ig^{-1} \d g  \Del_\mm 
e^{\Box_+ /M^2} \Del_\pd \frac{1}{\Box_+}\Delp
- \d \G_\mm   e^{\Box_+ /M^2}\Delpd  \frac{1}{\Box_+}\Delp 
\right) \d^{(4)}(z-z') \right] \nonumber \\
&=& -\int d^4z {\rm Tr} \left[ \L e^{\Box_+ /M^2}
\frac{1}{\Box_+}
\Delpd \Delp \Delmm \right] ~+~ {\rm h.c.}
\nonumber \\
&=& \int d^2x d \q^+ {\rm Tr} \left[ 
\L  e^{\Box_+ /M^2} \frac{1}{\Box_+}
\Delpd \Delp \Delmm \Delpd \d^{(4)}(z-z') \right] ~+~ {\rm h.c.}
\nonumber\\
&=&  i\int d^2x d\q^+ {\rm Tr} \left[ \L  e^{\Box_+ /M^2} 
\Delpd \d^{(4)}(z-z') \right] ~+~ {\rm h.c.} ~~~.
\ena
{}From the explicit expression (\ref{boxscalar}) for $\Box_+$ one
 realizes that in the limit $z' \to z$ the only nonvanishing 
contribution comes from the $W_- \Delp$ term in the exponential and the 
final result for the covariant anomaly is 
\beq
\d {\mit \G}_{cov} ~=~ - \frac{i}{4 \pi} \int d^2x d \q^+ ~{\rm Tr}
\left( \L W_-\right) ~+~ {\rm h.c.} ~~~.
\eeq
This expression only differs by an overall sign from the chiral anomaly 
(\ref{s1}) for the spinor case. Therefore, in the $(2,2)$ theory they 
cancel, consistent with the absence of anomalies  for that 
theory.

\subsubsection{The Consistency Terms}

~~~~To evaluate the consistency terms we use the following identities
for the homotopically extended derivative operators
\bea
\d_g \Del_+ &=& [\, \Del_+ ~,~ g^{-1} \d g \, ] ~~~,\nonumber \\
\d_g \Delp g^{-1} &=& - g^{-1} \d g \Delp g^{-1} ~~~, \nonumber \\
\d_g \Box_+ &=& -i \Del_\pd \,[\,\Del_+ ~,~ g^{-1} \d g \, ] 
\Tilde{\Del}_\mm ~-~i[\, \Delp ~,~ g^{-1} \d g \, ] \Tilde{\Del}_\mm 
\Del_\pd ~~~, \nonumber \\
\d_{\Tilde{\G}_\mm} \Tilde{\Del}_\mm &=& -i \d \Tilde{\G}_\mm  ~~~, 
\nonumber \\
\d_{\Tilde{\G}_\mm} \Box_+ &=& -\, \Del_\pd \Delp \d \Tilde{\G}_\mm ~-~ 
\Del_+ \d \Tilde{\G}_\mm \Del_\pd ~~~.
\ena
Therefore, inserting the regularized expression (\ref{scalarpropreg})
in (\ref{second}) and evaluating the gauge variations we have
(we omit an overall $\int_0^1 dy \int d^4z
\lim_{z' \to z} $ and a $\d^4(z-z')$)
\bea
&& \int d^4z'' \d g(z'') \frac{\d}{\d g(z'')} \VEV{ \frac{\d S}{\d g(z)}} =
-i \int_{\frac{1}{M^2}}^{\infty} dt
\Tilde{\Del}_\mm \Delpd e^{\Box_+ t} g^{-1} \d g \Delp g^{-1} 
\nonumber\\
&&~~~~~~~~+\int_{\frac{1}{M^2}}^{\infty} dt
\int_0^t ds \Tilde{\Del}_\mm \Delpd e^{\Box_+ s} [\Delp , g^{-1} \d g] 
\Tilde{\Del}_\mm \Delpd
e^{\Box_+  (t-s)} \Delp g^{-1}  \nonumber \\
&& \int d^4z''\d \Tilde{\G}_\mm(z'') \frac{\d}{\d \Tilde{\G}_\mm(z'')} 
\VEV{ \frac{\d S}{\d g(z)}}
= \int dt \d \Tilde{\G}_\mm \Delpd e^{\Box_+ t}
\Delp g^{-1}  \nonumber \\
&&~~~~~~~~
- i \int dt \int_0^t ds \Tilde{\Del}_\mm \Delpd e^{\Box_+ s} \Delp 
\d \Tilde{\G}_\mm \Delpd e^{\Box_+ (t-s)} \Delp g^{-1}
\nonumber \\
&& \int d^4z'' \d \Tilde{\G}_\mm(z'') \frac{\d}{\d \Tilde{\G}_\mm(z'')} 
\VEV{ \frac{\d S}{\d \Tilde{\G}_\mm(z)}} =
-\int dt \int_0^t ds \Delpd
e^{\Box_+ s} \Delp \d \Tilde{\G}_\mm \Delpd e^{\Box_+ (t-s)}
\Delp \nonumber \\
&& \int d^4z'' \d g(z'') \frac{\d}{\d g(z'')} \VEV{ \frac{\d S}{\d \Tilde{\G}_\mm(z)}}
=
\int dt \Delpd e^{\Box_+ t} [\Delp , g^{-1} \d g] \nonumber \\
&&~~~~~~~~
-i \int dt \int_0^t ds \Delpd e^{\Box_+ s} [\Delp , g^{-1} \d g]
\Tilde{\Del}_\mm \Delpd e^{\Box_+ (t-s)} \Delp ~~~.
\eea

Writing everything down one generates a large number of terms.  Using
the 
cyclicity of the trace and writing them somewhat symbolically, after
many
cancellations we have
\bea
-g^{-1} \pa_y g 
&[& i \Tilde{\Del}_\mm \Delpd  \eb  g^{-1} \d g \Delp + \Delp \d
\Tilde{\G}_\mm \Delpd \eb +
i\Delp g^{-1} \d g \Tilde{\Del}_\mm \Delpd \eb  \nonumber \\
&& + \Tilde{\Del}_\mm \Delpd \eb g^{-1}\d g \Delp \Tilde{\Del}_\mm 
\Delpd \eb \Delp -i\Delp
\Tilde{\Del}_\mm \Delpd \eb \Delp \d \Tilde{\G}_\mm \Delpd \eb \nonumber
\\
&&+ \Delp \Tilde{\Del}_\mm \Delpd \eb \Delp
g^{-1} \d g \Tilde{\Del}_\mm \Delpd \eb] \nonumber \\
- \pa_y \Tilde{\G}_\mm &[& +\Delpd \eb g^{-1} \d g \Delp -i\Delpd \eb
g^{-1} 
\d g \Delp \Tilde{\Del}_\mm \Delpd \eb \Delp] ~~~.
\eea
In this expression integration over $s$ and $t$ is understood, as above, and
the exponentials are also  essentially as above.

We now manipulate the double integral terms (those with two factors of
$\eb$
i.e. $ \cdots e^{\Box_+ s} \cdots e^{\Box_+ (t-s)} \cdots$) in the usual
fashion by writing  
\beq
\Delp \Tilde{\Del}_\mm \Delpd e^{\Box_+ (t-s)} \Delp ~=~
i \, \frac{\pa}{\pa t} \, e^{\Box_- (t-s)} \Delp ~~~,
\eeq
where $\Box_-$ has been defined in (\ref{scalarboxm}).  Performing
integration  by parts on $t$, we then obtain
\bea
&\int _0^{1/M^2} ds& \left[\, i (g^{-1} \pa_y g) \Tilde{\Del}_\mm
\Delpd 
e^{\Box_+ s} (g^{-1} \d g) \, e^{\Box_- ( 1/M^2 -s)} \Delp \right. 
\nonumber \\
&&\left. +i (g^{-1} \pa_y g)   \, e^{\Box_- (1/M^2-s)} \Delp (g^{-1} \d
g) 
\Tilde{\Del}_\mm \Delpd e^{\Box_+ s} \right. \\
&&\left. + (g^{-1} \pa_y g)\, e^{\Box_- (1/M^2-s)} \Delp \d
\Tilde{\G}_\mm
\Delpd e^{\Box_+ s} \right. \nonumber \\
&&\left. +  \pa_y \Tilde{\G}_\mm \Delpd e^{\Box_+ s} (g^{-1} \d g) 
\, e^{\Box_- (1/M^2-s)} \, \right] ~~~, \nonumber
\eea
leading to, in the usual manner,
\beq
\d {\mit \G}_{cons} = \frac{1}{4 \pi} \int_0^1 d y [ \, 2i (g^{-1} \d g)
\Tilde{\Del}_\mm (g^{-1} \pa_{y} g) + (g^{-1} \pa_{y} g) \d
\Tilde{\G}_\mm
- \pa_{y} \Tilde{\G}_\mm (g^{-1} \d g) \,] ~~~.
\eeq
Again, this expression differs from the spinor one (\ref{s2}) and 
(\ref{s3s4}) by an overall sign so that in the $(2,2)$ theory they
exactly cancel. Summing the covariant and the consistent terms the gauge 
anomaly in the chiral scalar case is then given by minus the result in 
(\ref{BGJanomaly}).

\section{Reduction to Components in WZ Gauge}

~~~~In ref. \cite{GGP} we showed that the supersymmetric anomaly (\ref{BGJ1})
contains the correct bosonic component by reducing
it to components in the Wess--Zumino gauge.
In this section we show that the  anomaly
(\ref{BGJanomaly}) reduces correctly to the corresponding component result.

In the bosonic case, the 2D Yang-Mills covariant derivative is defined
in terms of a two--components vector $(A_\pp,A_\mm)$ as
\bea
&& \Del_\pp = \pa_\pp ~-~ iA_\pp  \quad , \quad
\Del_\mm ~=~ \pa_\mm ~-~ iA_\mm ~~~, \nonumber \\
&& [ \Del_\pp \, , \, \Del_\mm ] ~=~ -i F_{\pp \mm} ~=~
-i( \, \pa_\pp A_\mm ~-~ \pa_\mm A_\pp ~-~ i[A_\pp \, , \, A_\mm] \, )
~~~.
\label{comp2}
 \ena
We consider a theory of chiral fermions minimally coupled to a set of 
gauge fields
\beq
S ~=~ i \int d^2x \, \bar{\zeta}_\md  \Del_\pp \zeta_- ~~~.
\eeq
This action is invariant under the gauge transformations
\bea
&& \d_G(\l) \zeta_- ~=~ i \l \zeta_-  \qquad , \qquad  
\d_G(\l) \bar{\zeta}_\md ~=~ -i \bar{\zeta}_\md \l ~~~,
\nonumber \\
&& \d_G(\l) A_\pp ~=~ \pa_\pp \l ~+~ i[\l, A_\pp] ~~~.
\label{gaugetranfs}
\ena
Following the discussion in \cite{GGP}, we compute algebraically the  
2D consistent anomaly by directly solving the Wess--Zumino
consistency condition
\beq
\d_G(\l_1) {\cal A}_{\rm BGJ}(\l_2) ~-~ \d_G(\l_2) {\cal A}_{\rm
BGJ}(\l_1)
~=~ -i {\cal A}_{\rm BGJ} ([\l_1,\l_2]) ~~~.
\label{WZ}
\eeq
To this end, we consider a set of basis monomials defined as
\beq
{\cal M}_0 (\l) ~=~ \frac12 \e^{\underline{a} \, \underline{b}}~ {\rm
{Tr}}
\Big\{ \l \,  F_{\underline{a} \, \underline{b}} \, \Big\}  ~~~, ~~~
{\cal M}_2 (\l) ~=~  \e^{\underline{a} \, \underline{b}} ~ {\rm {Tr}}
\Big\{\l \,A_{\underline{a}} A_{\underline{b}}\, \Big\} ~~~.
\eeq
In terms of light--cone coordinates they read
\bea
{\cal M}_0 (\l) &&=~ {\rm {Tr}} \Big\{ \l \,  F_{\pp \, \mm} \,
\Big\} ~=~ {\rm {Tr}} \Big\{ \l \, (\, \pa_{\pp} A_{\mm} \,-\,
\pa_{\mm} A_{\pp} \,-\,  i \, [ A_{\pp} , A_{\mm} ] \, )~
\Big\} ~~~, \nonumber \\
{\cal  M}_2 (\l) &&=~  {\rm {Tr}} \Big\{\l
\,(\,A_{\pp} A_{\mm}\, -\,A_{\mm} A_{\pp}\, )\, \Big\} ~~~.
\ena
We look for a solution of the WZ  consistency condition as linear
combination 
of the two previous monomials. Imposing the condition (\ref{WZ}) on the 
general structure
\beq
c_0 {\cal M}_0 (\l) ~+~ c_2 {\cal M}_2(\l) ~~~,
\eeq
we find a solution for $c_2 = i c_0$. Therefore, this leads to the
identification
\bea
{\cal A}_{\rm {BGJ}}(\l)  &\equiv& C_0 \,
\int d^2x  ~ \left( {\cal M}_0 ~+~ i {\cal M}_2 \right)
\nonumber \\
&=& C_0 \, \int d^2x \, {\rm {Tr}} \Big\{ \l \, (\, \pa_{\pp}
A_{\mm} \,-\,  \pa_{\mm} A_{\pp}  \, )~  \Big\} ~~~,
\label{bosanomaly}
\ena
with $C_0$ an overall normalization. The comparison with the
perturbative 
calculation gives $C_0 = - (1/ 8 \p)$. In 2D, the consistent anomaly has 
the same form as the abelian anomaly.

The consistent anomaly is always defined up to the variation of
a local functional. In the two dimensional case, an alternative
expression for the anomaly is
\beq
{\cal A}_{\rm BGJ} (\l) ~=~ -2C_0 \int d^2x \, {\rm Tr} \Big\{ \,
\l \, \pa_\mm A_\pp \, \Big\} ~~~,
\label{bosanomaly2}
\eeq
which differs from (\ref{bosanomaly}) by the variation 
\beq
\d \int d^2x \, {\rm Tr} (A_\pp A_\mm) ~=~
- \int d^2x \, {\rm Tr} \left[\, \l \, \left(
\pa_\pp A_\mm ~+~ \pa_\mm A_\pp \right) \, \right] ~~~.
\label{trivial}
\eeq

We now consider the supersymmetric expression (\ref{BGJanomaly}) 
for the $(2,0)$ anomaly written as a sum of four contributions
\beq
{\cal A}_{\rm BGJ} = (\frac{1}{4 \pi}) \, 
( {\cal A}_1 + {\cal A}_2 +
{\cal A}_3 + {\cal A}_4) ~~~,
\label{sum}
\eeq
where we have defined
\bea
&& {\cal A}_1 ~=~
i \int d^2x d \theta^+ {\rm Tr} ( \L W_-) + {\rm h.c.} ~~~, \nonumber \\
&& {\cal A}_2 ~=~
-2 i \int d^2x d^2 \theta \int_0^1 d y \, {\rm Tr} \left[ (g^{-1}
\d g) \Tilde{\Del}_\mm ( g^{-1} \pa_{y} g) \right] ~~~, \nonumber \\
&& {\cal A}_3 ~=~ -\int d^2x d^2 \theta \int_0^1 d y \, {\rm Tr} \left[
(g^{-1} \pa_{y} g) \d \tilde{\G}_\mm \right] ~~~, \nonumber \\
&& {\cal A}_4 ~=~ \int d^2x d^2 \theta \int_0^1 d y \, {\rm Tr} \left[
(g^{-1} \d g) \pa_{y} \tilde{\G}_\mm \right] ~~~,
\label{4terms}
\eea
and we perform the reduction separately for the four terms.

The WZ gauge is defined by $V| = D_+ V| = 0 $. In this gauge the bosonic
components are
\bea
&& \G_\pp | ~=~ D_\pd D_+ V| ~=~ A_\pp ~~~, \nonumber \\
&& \G_\mm | ~=~ A_\mm ~~~, \nonumber \\
&& [\, \Delpp ~,~ \Delmm \, ] | ~=~ ( \Delpd \Bar{W}_{\md} \,-\, \Delp
W_- )|
= -i F_{\pp \mm} ~~~.
\label{comp1}
\eea
with $F_{\pp \mm}$ as above.
Moreover, using the conditions defining the WZ gauge, the chirality of
$\L$
and   the identity $\L| = \bar{\L}| = \l$, one finds
\bea
&& g^{-1} \d g\Big| ~=~=
i y \, g^{-1} \Big[ \, \bar{\L} e^V ~-~ e^V \L \,
\Big]\Big| ~=~ 0 ~~~, \nonumber \\
&& g^{-1} \pa_{y} g\Big| ~=~ g^{-1} (e^V -1)\Big| ~=~ 0 ~~~, \nonumber
\\
&& D_+ D_\pd \, ( g^{-1} \d g)\Big| ~=~
-y \, \left( \pa_\pp \l ~-~ i[ A_\pp \, , \, \l] \right)
~=~ -y \, \Del_\pp \l ~~~.
\label{reduction}
\ena

In order to perform the reduction to components we need to choose a
particular
function $\Tilde{\G}_\mm$. Some indications on the particular structure
of 
this function can be obtained by defining the extended $(2,0)$ SYM
theory
as the reduction of an extended $(2,2)$ SYM theory. In fact, in the
latter 
theory the supersymmetry algebra constrains the particular dependence on
$g$
of  both vector connections in $\Del_\pp$ and $\Del_\mm$, once the 
homotopic path has been fixed in the definition of the extended
spinorial 
derivatives (see eq. (\ref{extender})).  If we call ${\cal H}$ the
homotopy
path of the $(2,2)$ theory (${\cal H}|_{\theta_- = \theta_\md =0} = g$),
in
chiral representation the extended $(2,0)$ connection is obtained from
the
$(2,2)$ one, as an extension of (\ref{gamma}) 
\beq
\Tilde{\G}_\mm ~=~ D_\md ( {\cal H}^{-1} D_- {\cal H})\Big|_{\theta^- = 
\theta^\md =0} ~~~.
\eeq
Computing the gauge variation and the $y$--derivative in the $(2,2)$
theory and performing the complete reduction to the component theory,
for the bosonic components we find 
\bea
&& \Tilde{\G}_\mm \Big| ~=~ y A_\mm ~=~ y \G_\mm \Big| ~~~,
\nonumber \\
&& \d \Tilde{\G}_\mm \Big| ~=~ y \pa_\mm \l
- iy [A_\mm, \l] ~=~ y \Del_\mm \l ~=~ y ~\d \G_\mm \Big| ~~~,
\nonumber \\
&& \pa_{y} \tilde{\G}_\mm \Big|  ~=~ A_\mm ~=~ \G_\mm \Big| ~~~.
\label{reduction2}
\ena
The previous identities strongly suggest the following general structure
for the homotopic extension of $\G_\mm$ directly in $(2,0)$ superspace
\beq
\Tilde{\G}_\mm ~=~ f(y) \G_\mm ~+~ \Delta \G_\mm ~~,
\label{general}
\eeq
where $f(0) =0$ and $f(1)=1$, consistently with the choice of the
boundary
conditions for $g$, and $\Delta \G_\mm| =0$.
We note that for $f(y) =y$ we obtain the extension coming from
the $(2,2)$ theory, whereas $f(y) =1$ and $\Delta \G_\mm =0$ 
corresponds to no extension for $\G_\mm$.  

We perform the reduction of (\ref{BGJanomaly}) to the bosonic components
by
assuming for $\Tilde{\G}_\mm$ the form (\ref{general}).  We first
consider 
the covariant term $\tilde{\cal A}_1$. 
Keeping only the bosonic components we have
\beq
\eqalign{ {~~~~~~~~~~~~}
{\cal A}_1 &=~ i\int d^2x ~{\rm Tr} ( \L D_+ W_-) |~+~ {\rm h.c.} ~=~
- \int d^2x ~ {\rm Tr} (\l F_{\pp \mm}) {~~~~~} \cr
&= ~\int d^2x ~{\rm Tr} \left\{\,  \l \left( \pa_\mm A_\pp - \pa_\pp
A_\mm + i [ A_\pp , A_\mm ] \, \right) \,\right\} ~~~,
}\label{an1}
\eeq
where we have used $\L| = \bar{\L}| = \l$ and eq. (\ref{comp2}).

For the first consistent term ${\cal A}_2$, using the identities in 
(\ref{reduction}) we find
\beq
{\cal A}_2 = 
-2i \int d^2x  \int_0^1 d y y \, D_+ D_\pd {\rm Tr} \left[
g^{-1} \d g \tilde{\Del}_\mm ( g^{-1} \pa_{y} g) \right]
\Big| ~\rightarrow~ 0 ~~~.
\eeq
Therefore it never contributes to the bosonic anomaly, independent
of the choice of the homotopic path.

Using eq. (\ref{reduction2}) the ${\cal A}_3$ term gives (again
neglecting fermionic terms)
\bea
{\cal A}_3 &&= -\int d^2x \int_0^1 dy \, D_+ D_\pd {\rm Tr} \left[\, 
g^{-1} (e^V - 1) \, \d \tilde{\G}_\mm \right] \Big| \nonumber \\
&& {~~}\rightarrow -\int d^2x \int_0^1 dy \, {\rm Tr} \left[\, 
(D_+ D_\pd V) \d \tilde{\G}_\mm \right] \Big| \nonumber \\
&& =  \int d^2x \int_0^1 dy \, f(y) \, {\rm Tr} \{ A_\pp ( \pa_\mm \l -
i [ A_\mm , \l ]) \} \nonumber \\
&& = \int_0^1 dy f(y) ~ \int d^2x \, {\rm Tr} \left\{ \l \Big( -\pa_\mm 
A_\pp ~-~ i  [ A_\pp ,A_\mm ] \Big) \right\} ~~~.
\label{an3}
\eea

Finally, using eqs. (\ref{reduction},\ref{reduction2}) in the last
consistent term ${\cal A}_4$ we have
\bea
{\cal A}_4 &&= \int d^2x \int_0^1 dy \, D_+ D_\pd {\rm Tr} \left[ \,
g^{-1}
\d g  ~\pa_{y} \tilde{\G}_\mm \right] \Big| \nonumber \\
&& {~~}\rightarrow \int d^2x \int_0^1 dy \, {\rm Tr} \left[\, [D_+ D_\pd
(g^{-1} \d g)] \pa_{y} \tilde{\G}_\mm \right] \Big|
\nonumber \\
&& =~ \int d^2x \int_0^1 dy \, y f'(y) \, {\rm Tr} \{ \,( -\pa_\pp \l +
i [ A_\pp , \l ]) A_\mm \}
\nonumber \\
&& =~ \left( 1 - \int_0^1 dy f(y) \right) \,\int d^2x \, {\rm Tr}
\left\{ 
\l \Big( \pa_\pp A_\mm ~-~ i [ A_\pp ,A_\mm ] \Big) \right\} ~~~.
\label{an4}
\eea
Summing the three nonvanishing contributions it is easy to see that the
terms proportional to $[ A_\pp , A_\mm]$ cancel independent of the
choice of $f$ and we are left with
\beq
{\cal A}_{\rm BGJ}^{bos}(\l) ~=~ \frac{1}{4\pi} \int d^2x ~{\rm Tr} 
\left\{ \l \left[\, \pa_\mm A_\pp  ~-~ \int_0^1 dy f(y) ~ (\pa_\mm A_\pp 
+ \pa_\pp A_\mm)\, \right] \right\} ~~~.
\label{anomaly2}
\eeq
According to eq.(\ref{trivial}) the second term is cohomologically
trivial 
and can be neglected in the final expression for the anomaly which turns 
out to be in agreement with (\ref{bosanomaly2}) (or equivalently
(\ref{bosanomaly})). 

{}From the above result we can conclude that the choice of a particular 
function $f$ affects the final answer only by terms which are the 
variation of a local functional.   Therefore, we are allowed to choose 
in (\ref{general}) {\em any} regular function in $[0,1]$ with correct 
boundary conditions.  The most general homotopically  extended $(2,0)$
theory depends  on the choice of two independent paths $g$ and
$\Tilde{\G}_\mm$.  
In particular, one could choose $f(y)=1$ which corresponds to performing 
the entire  derivation of the supersymmetric anomaly without
homotopically
extending $\G_\mm$.  This might be {\em a priori} expected in the case 
of a chiral spinor superfield since $\G_\mm$ never couples to the
spinor. Instead in the case of a chiral scalar it is not an obvious
result.

Another choice that may ultimately prove of use in other contexts is
$f(y)
=y$ for which we find agreement between the perturbative result and the 
solution of the 2D WZ consistency condition given in (\ref{bosanomaly}).
This choice of bosonic homotopy for the minus--minus connection
resembles 
the minimal supersymmetric homotopy (\ref{choicemin}) in that it
satisfies
equations analogous to (\ref{NwTN2}) and is the one obtained by the 
truncation described in (\ref{reduction2}).   
For this choice, utilizing the minimal supersymmetric homotopy 
implies that we can re-write the last three terms in (\ref{sum}) 
and thus express the {\it {holomorphic}} consistent (2,0) Yang-Mills 
anomaly, as in ref. \cite{GGP},
in the form of the imaginary part of a superaction, ${\cal 
A}_{\rm BGJ} = {\cal {I}{\rm m}} [{\Tilde {\cal  A}}{}_{\rm BGJ} \,]$ 
where
\beq
\eqalign{
{\Tilde {\cal A}}{}_{\rm BGJ} &=~  -\, ({ 1 \over 2 \p}) \,   \int 
d^2 x \, d^2 \q ~ {\cal P}_{\mm} ( \L\,; e^V) ~~~, 
{~~~~~~~} 
} \label{4terms4}\eeq
with the 2D, $N$ = (2,0) ${\cal P}$-superfunction given by
\beq
\eqalign{ 
{\cal P}_{\mm}( \L\,; e^V) = &~ {\rm Tr} \Big\{ \, \L 
\Big( \, \G_{\mm} \,-\, \int_0^1 d y \,y ~ \Big[ \, 
\G_{\mm}\, e^V {\cal G} \, +\, [\, {\G}{}_{\mm} ~,~ {\cal G} \,
(e^V - 1\,) \,] \,  (\, 1 ~-~ 2y \, e^V \, {\cal G} \,)  
{~~}  \cr 
&{~~~~~~~~~~~~~~~~~~~~~~~~~~~~~}+~ i \,[\, {\pa}{
}_{\mm} ({\cal G} \,(e^V - 1\,)) \, ] \, (\, 1 - 2 
e^V \, {\cal G} ) ~ \Big] ~ \Big)  ~ \Big\} ~~~. }
\eeq
where again ${\cal G} = [1+y(e^V-1)]^{-1}$.

\sect{Conclusions and Summary}

~~~~In this paper, we have presented examples of the construction of the
superspace consistent anomaly using the method of Banerjee, Banerjee and
Mitra \cite{BAN}. We have shown that the
 4D, $N$ = 1 supersymmetric extension 
of the consistent anomaly has a preferred choice of homotopy which 
leads to vast simplification in the form of the BGJ non-Abelian 4D, 
$N$ = 1 consistent anomaly action.  We believe  that this choice, which we
call the minimal homotopy \cite{GGP}, yields substantial clarity to the issue of
4D, $N$ = 1 supersymmetric anomalies over previous results in the
literature \cite{PIG}-\cite{OHS}.  We have also shown that any
other choice of homotopy utilized to write the 4D, $N$ = 1
supersymmetric BGJ action yields a non-minimal form of the anomaly 
that differs from the minimal one only by the gauge variation of a 
local counterterm.  This provides a  proof of a previous 
assertion \cite{GGP} that homotopic non-minimality was responsible 
for the opacity of most discussions of this topic.  Our discussion 
has also pointed out the differential equation (\ref{NwTN2}) that 
singles out the 
minimal homotopy from all others.  This provides a finer definition 
of the minimal homotopy.

Turning to the issue of the universality of the role of the minimal
homotopy in supersymmetric gauge theories, we have explored this
outside of 4D, $N$ = 1 theories.  For 2D, $N$ = (2,2) models,
involving only chiral matter coupled to ordinary vector multiplets,
all such anomalies are found to vanish as expected from the fact
that these are vector-like theories.  
%This leaves open slightly
%the issue of whether this also holds for theories involving twisted 
%matter and gauge multiplets as well as those involving their non-minimal 
%counterparts (and their twisted versions).  In relation to this set of issues 
%is also the question of the impact of mirror symmetry?  These
%questions might provide the basis for future investigations.

For 2D, $N$ = (2,0) models, anomalies are expected and our study
has determined their form.  In this context, the minimal homotopy
was also found to lead to a simple expression.  We also found that 
the (2,0) reduction of
the homotopic extension of the 2D, $N$ = (2,2)  (the analog of
(2.5)) naturally leads to the use of the minimal homotopical
extension for the connection in the non-supersymmetric sector of
the theory.

We end this by again noting the fact that the existence of
the minimal homotopy obeying (2.30), playing a special role
in 4D, $N$ = 1 theories,  raises a question of whether
the five-dimensional embedding space ($x$, $y$) may be playing
a more subtle role than may have  been suspected.

\newpage
\appendix

\sect{Four Dimensional $N=1$ Superspace Conventions} 

~~~~In four dimensions we use {\em Superspace} \cite{GAT}
notations and conventions
supplemented by the conjugation rule $(D_\a)^{*} = - \bar{D}_{\dot\a}$,
$(D^\a)^{*} = \bar{D}^{\dot\a}$ and $(\pa_a)^{*} 
= \pa_a$ ($a = \a \ad$).

In chiral representation, the superspace Yang--Mills covariant
derivatives 
$\Del_{\underline{A}} \equiv (\Del_\a, \Del_{\dot\a} ,
\Del_a)$ 
with $\Del_{\underline{A}} = D_{\underline{A}} - i \G_{\underline{A}}$
are
defined in terms of a real vector superfield $V$ as
\beq
\Del_\a ~\equiv~ e^{-V} D_\a e^V \quad , \quad \Del_{\dot\a} 
~\equiv~ \bar{D}_{\dot\a}  \quad , \quad \Del_a 
~\equiv~ -i \,\{\, \Del_\a ~,~ \Del_{\dot\a} \, \} ~~~.
\label{4derivatives}
\eeq
The corresponding spinorial field strengths are (the field
strength $\Bar{W}_{\dot\a}$ below is the conjugate of $W_\a$; note that
$\bar{\Gamma}_\md$ should  not be confused with $\Gamma_\md$), 
\beq
W_\a ~=~ \bar{D}^2 \G_\a =~=~ i \bar{D}^2 (e^{-V} D_\a e^V) \quad ,
\quad \Bar{W}_{\dot\a} ~=~ D^2 \bar{\G}_{\dot\a} ~=~ i D^2(e^V
\bar{D}_{\dot\a} e^{-V}) ~~~.
\label{4field}
\eeq

The infinitesimal gauge transformations are generated by  chiral 
and antichiral superfield parameters $\L$, $\bar{\L}$ ($\bar{D}_{
\dot\a} \L =0$, $D_\a \bar{\L} =0$)  given by
\bea
&& \d e^V ~=~ i \,[\, \bar{\L} e^V ~-~ e^V \L \,] \quad , \quad
\d e^{-V} ~=~ i \,[\, \L e^{-V} ~-~ e^{-V} \bar{\L} \,] ~~~, 
\nonumber \\
&& \d \G_{\underline{A}} ~=~ D_{\underline{A}} \L ~+~ i 
[ \L , \G_{\underline{A}} ] \quad , \quad \d W_\a ~=~ i [\, \L ~,~ 
W_\a \,] ~~~.
\label{gaugetr}
\ena
The action describing scalar matter coupled to Yang--Mills is
\beq
S ~=~ \int d^8z ~ \bar{\Phi} e^V \Phi ~~~, 
\label{4daction}
\eeq
where $\bar{D}_{\dot\a} \Phi = D_\a \bar{\Phi} =0$. It is invariant
under gauge transformations (\ref{gaugetr}) supplemented by
\beq
\d \Phi ~=~ i\L \Phi \qquad , \qquad
\d \bar{\Phi} ~=~ -i \bar{\Phi} \L  ~~~.
\eeq

\sect{$N=2$ Superspace and Superfields in Two Dimensions}

\subsection{$(2,2)$ Superspace}

~~~~The two--dimensional superspace description of a $(2,2)$ theory is
essentially the same as $N=1$ in four dimensions with a suitable 
identification of the spinorial coordinates.  It is described in terms
of
two light--cone coordinates $(x^\pp,x^\mm)$ and four spinor coordinates
$(\theta^+, \theta^\pd, \theta^-, \theta^\md)$ where plus and minus
label the two chiral sectors. 

An untwisted scalar $(2,2)$ supersymmetric theory is described by the
superspace action
\beq
S ~=~ \int d^2x d^4 \theta \, \bar{\bf \Phi} {\bf \Phi}  ~~~, 
\label{22action}
\eeq
where ${\bf \Phi}$ and $\bar{\bf \Phi}$ are chiral and antichiral fields 
respectively ($D_+ \bar{\bf \Phi} = D_- \bar{\bf \Phi} = D_\pd {\bf
\Phi} = D_\md {\bf \Phi} =0$)
and
\beq
d^4 \theta ~\equiv~ d\theta^+ d\theta^\pd d\theta^- d\theta^\md ~\equiv~
D_+ D_\pd D_- D_\md ~~~.
\eeq
A twisted version of the theory is also consistent \cite{HPS,GGK} in 
terms of a twisted scalar superfield defined by $D_\pd {\bf \Phi} = D_-
{\bf \Phi} = 0$.

The minimal coupling to Yang--Mills fields is realized by introducing
covariant derivatives.  Also in this case there exist two alternative 
formulations \cite{HPS,GGK} in terms of Yang--Mills multiplets and 
twisted Yang--Mills multiplets, respectively. However,  we consider only
the version described by the constraints 
 $$ \eqalign{
 \{\, \nabla_+ ~,~ \nabla_+ \,\} ~=~ 0 ~~~&,~~~ \{\, \nabla_- ~,~ 
\nabla_- \,\} ~=~ 0 ~~~, \cr
\{\, \nabla_+ ~,~ { \nabla}_- \,\} ~=~ 0~~~&, ~~~ \{\, \nabla_+ ~,~ {\nabla_{
\Dot -}} \,\}  ~=~ i {\Bar { W}}~~~, \cr
\{\, \nabla_\Dot + ~,~ { \nabla}_\Dot - \,\} ~=~0~~~&, ~~~ \{\, \nabla_\Dot + ~,~ 
{\nabla_{-}} \,\}  ~=~ -i { W}{} ~~~, \cr
\{\, \nabla_+ ~,~ {\nabla}_{\Dot +} \,\} ~=~ i  \nabla_{\pp} 
~~~&,~~~ \{\, \nabla_- ~,~ {\nabla}_{\Dot -} \,\} ~=~ i  \nabla_{
\mm}  ~~~, \cr
[\, \nabla_+ ~,~ \nabla_{\pp} \,] ~=~ 0 ~~~&,~~~ \{\, \nabla_- 
~,~ \nabla_{\mm} \,\}  ~=~ 0    ~~~, }$$
$$ [\, \nabla_+ ~,~ \nabla_{\mm} \,] ~=~ -\,   {\nabla}_- 
{\Bar { W}}   ~~~, $$
$$ \{\, \nabla_- ~,~ \nabla_{\pp} \,\} ~=~ +   {\nabla}_+ {{ 
W}} ~~~, $$
\beq [\, \nabla_{\pp} ~,~ \nabla_{\mm} \,] ~=~ - i {\cal F}{}~~~.
{~~~~}
\eeq
The superfields ${ W}$ and  ${\cal
F} $  satisfy the identities
$$
\nabla_{-} { W}{} ~=~ \nabla_{\Dot +} { W}{}~=~ 0 ~~~,
$$
\beq
{\cal F} ~=~  [\,  \nabla_{+} \nabla_{\Dot -} {{ W}}
 ~-~ \nabla_{\Dot +} \nabla_{-} {\Bar { W}}  
\,]  ~~~.
\eeq
The solution to the constraints, as in 4D, can be expressed in terms of the
hermitean prepotential ${\cal V}$.

\subsection{$(2,0)$ Superspace}

~~~~The chiral $(2,0)$ ($(0,2)$) theory can be obtained as a reduction
of the $(2,2)$ theory
by setting $\theta^- = \theta^\md =0$ ($\theta^+ = \theta^\pd =0$).

{}From the action (\ref{22action}), by explicitly integrating on
$(\theta^-, \theta^\md)$ one obtains
\beq
S ~=~ \int d^2x d\theta^+ d\theta^\pd \, \left[ - D_\md \bar{\bf \Phi}
D_- {\bf \Phi} ~-~ i\bar{\bf \Phi} \pa_\mm {\bf \Phi} \right] 
\Big|_{\theta^-=\theta^\md =0} ~~~, 
\label{20actions}
\eeq
where the chirality constraints have been used. The two terms represent 
the actions for a chiral $(2,0)$ spinor superfield $\chi_- \equiv D_-
{\bf \Phi}
|_{\theta^- = \theta^\md =0}$ and a chiral $(2,0)$ scalar $\Phi \equiv
{\bf
\Phi}|_{\theta^- = \theta^\md =0}$, respectively.

In the $(2,0)$ sector we use the following conventions for functional
derivatives. For the  chiral scalar superfield we define
\beq
\frac{\d{~~~~~}}{\d \Phi (z')} \int d^2x d \theta^+ f( \Phi (z)) ~=~ 
f'(\Phi(z')) ~~~,
\eeq
\beq
\frac{\d{~~~~~}}{\d \bar{\Phi} (z')} \int d^2x d \theta^{\pd} f( 
\bar{\Phi} (z)) ~=~ f'(\bar{\Phi}(z')) ~~~.
\label{functder}
\eeq
It follows that the functional derivatives are
\beq
\frac{\d \Phi(z)}{\d \Phi(z')} = D_{\pd} \d^{(4)}(z-z')
\qquad ,\quad
\frac{\d \Phib(z)}{\d \Phib(z')} = -D_+ \d^{(4)}(z-z') ~~~,
\label{der}
\eeq
where the derivatives on the r.h.s. act on $z$ and we have defined
\beq
\d^{(4)}(z-z') ~\equiv ~ \d^{(2)}(x-x') \, \d^{(2)}(\q -\q') ~\equiv ~
\d^{(2)}(x-x') \, (\q^{\pd} -{\q^{\pd}}')
 (\q^+  -{\q^+}') ~~~.
\label{delta}
\eeq
Using these conventions the scalar propagator is given by
\beq
\langle \Phi(z) \Phib (z') \rangle \,\equiv\, \int D\Phi D \Phib \,
e^{-S} \Phi(z) \Phib (z') \,=\, - \frac{\d}{\d J(z)} \frac{\d}{\bar{
J}(z')} {\cal W}[J, \bar{J}] \Big|_{J = \bar{J} \,=\,0 ~~~,}
\label{definition}
\eeq
where
\beq
{\cal W}[J, \bar{J}] ~=~ \int D\Phi D \Phib \, \exp{\{-S - \int d^2x
d\theta^+ J \Phi + \int d^2x d \theta^\pd \Phib \bar{J} \, \} } ~~~.
\eeq
In the previous expression the sources $J$ and $\bar{J}$ are spinors
and anticommute with the measure.

In the case of the  chiral  spinor superfield, using the same definitions
(\ref{functder}) for the functional derivatives we obtain
\beq
\frac{\d \chi_-(z)}{\d \chi_-(z')} ~=~ -D_{\pd} \d^{(4)}(z-z')
\qquad ,\quad
\frac{\d \chib_{\md}(z)}{\d \chib_{\md}(z')} ~=~ D_+ \d^{(4)}(z-z')
~~~,
\label{der2}
\eeq
where the derivatives on the r.h.s. act on $z$.  The propagator is then
defined as
\beq
\langle \chi_-(z) \chib_{\md} (z') \rangle \,\equiv\, \int D\chi_- D
\chib_{\md} \, e^{-S} \chi_-(z) \chib_{\md} (z') \,=\, - \frac{\d}{\d 
J(z)} \frac{\d}{\d \bar{J}(z')} {\cal W}[J, \bar{J}] \Big|_{J=\bar{J}=0}
~~~,
\eeq
with
\beq
{\cal W}[J, \bar{J}] = \int [D \chi_- D \chib_{\md}] \exp{ \{ -S + \int 
d^2x d \theta^+ J \chi_- - \int d^2x d \theta^{\pd} \chib_{\md} \bar{J} 
\} } ~~~.
\eeq
and scalar sources.

The $(2,0)$ Yang--Mills theory can also be obtained as a reduction of
the
$(2,2)$ theory, setting $\theta^- = \theta^\md = 0$. 
In this case the constraints are
\beq
\eqalign{
[\, \Del_+ ~,~ \Del_+ \, \} &=~ [ \, \Del_\pd ~,~ \Del_\pd \, \} ~~~,
\cr
[\, \Del_+ ~,~ \Del_\pd \,\} &=~ i \, \Del_\pp ~~~, \cr
[\, \Del_+ \, , \, \Del_\mm \, \} &=~ i \, W_\md \qquad , \quad
[\, \Del_\pd ~,~ \Del_\mm \,\} ~=~ -i \, W_- ~~~, \cr
[\, \Del_\pp ~,~ \Del_\mm \,\} &=~ \Del_\pd W_\md ~-~ \Del_+ W_-
~~~. }
\label{20constraints}
\eeq
In chiral representation, they are solved by
\beq
\Del_+ = e^{-V} D_+ e^V \qquad ,\qquad \Del_\pd = D_\pd ~~~,
\eeq
where $V$ is the $(2,0)$ vector multiplet, $V \equiv {\cal V}|_{\theta^-
= \theta^\md =0}$.  From the previous relations it follows
\beq
\G_+ = i(e^{-V} D_+ e^V) \quad , \quad
\G_\pd = 0 \quad , \quad \G_\pp = -i D_\pd \G_+ ~~~,
\eeq
The connection $\G_\mm$ is an independent superfield.  However, if 
the theory is obtained by reduction of  the $(2,2)$ theory one finds
\beq
\G_\mm ~=~ D_\md (e^{-{\cal V}} D_- e^{\cal V}) \Big|_{\theta^-
~=~ \theta^\md ~=~0} ~~~.
\label{gamma}
\eeq
Moreover, one also finds
\bea
&& W_- ~=~ D_\pd \G_\mm ~~~, \nonumber \\
&& W_\md ~=~ \pa_\mm \G_+ ~-~ D_+ \G_\mm ~+~ i[\G_+, \G_\mm] ~=~
\pa_\mm \G_+ ~-~ \Del_+ \G_\mm ~~~.
\ena
These two superfields satisfy the chirality constraints $\Del_\pd W_- = 
\Del_+ W_\md =0$.

As described at the beginning of this section, possible $(2,0)$ matter 
fields are chiral scalars $\Phi$ and chiral spinors $\chi_-$.  The
relevant actions for matter coupled to gauge fields are then
\bea
&& S_{\Phi} ~=~ -i\int d^2x d^2\theta \bar{\Phi} e^V \Del_\mm \Phi
~=~ -i \int d^2x d^2\theta \bar{\Phi} e^V ( \pa_\mm - i\G_\mm) \Phi
~~~, 
\label{scalaract} 
\\
&& S_{\chi} ~=~ -\int d^2x d^2\theta \bar{\chi}_\md e^V \chi_- ~~~.
\label{spinoract}
\ena
They are invariant under gauge transformations
\bea
&& \Phi \rightarrow e^{i\L} \Phi \qquad , \qquad \bar{\Phi} \rightarrow
\bar{\Phi} e^{-i\bar{\L}} ~~~, \nonumber \\
&& \chi_- \rightarrow e^{i\L} \chi_- \qquad , \qquad \bar{\chi}_\md
\rightarrow \bar{\chi}_\md e^{-i\bar{\L}} ~~~, \nonumber \\
&& e^V \rightarrow e^{i\bar{\L}} e^V e^{-i\L} ~~~, \nonumber \\
&& \Del_\mm \rightarrow e^{i\L} \Del_\mm e^{-i\L} ~~~.
\label{gvariation}
\ena
The total action $S_{\Phi} + S_{\chi}$ gives $(2,2)$ matter coupled
to Yang--Mills fields.

\sect{Matter Propagators}

~~~~In this appendix we derive the expressions for  the exact propagators
of the  $N=1$ scalar
superfield in four dimensions coupled to gauge fields and for the
$(2,0)$
chiral scalar and spinor superfields coupled to 2D Yang--Mills fields.
{}From the four dimensional propagator we immediately read also the result
for a scalar multiplet coupled to Yang--Mills in the $(2,2)$ theory. 

\subsection{N=1 Scalar Superfield}

Here we review the calculation of the covariant propagator
for a chiral scalar minimally coupled to Yang--Mills in four dimensions. 
One way to find the propagator is to determine the Schwinger-Dyson
equation it satisfies \cite{HAY}. Taking into account the explicit
expression
(\ref{4daction}) for the scalar action, we write the functional integral
\beq
\VEV{\Phib(z)}= \int D \Phi D \Phib \exp[ - \int d^8w \Phib (w) e^V
\Phi (w) ] \cdot  \Phib (z) ~~~,
\eeq
and make a change of variable $\Phib \rightarrow  \Phib + \d \Phib$
under which the functional integral doesn't change. One gets then
\beq
0~=~\int D \Phi D \Phib \exp[ - \int d^8w \Phib (w) e^V \Phi (w) ] 
\cdot [-\int d^8u \d \Phib (u) e^V \Phi(u) \Phib (z) + \d \Phib (z) ]
~~~. \eeq
Now taking the functional derivative $\d / \d \Phib (z') $ and
reinterpreting the
functional integral as giving an expectation value, we obtain
\beq
\int d^8u D^2 \d^{(8)} (u-z')  \, e^V \VEV{\Phi(u) \Phib (z)}- 
D^2 \d^{(8)} (z-z')  ~=~ 0 ~~~,
\eeq
or
\beq
D^2 e^V \VEV{\Phi(z') \Phib (z)} ~=~ D^2 \d^{(8)} (z'-z) ~~~.
\eeq
Now, multiplying by $e^{-V}$ on the left we have
\beq
\Del^2 \VEV{\Phi(z') \Phib (z)} ~=~  e^{-V} D^2  \d^{(8)} (z'-z) ~=~
\Del^2 e^{-V} \d^{(8)} (z' - z) ~~~.
\eeq
We proceed then multiplying  by $\Bar{\Del}^2$ and extending $\Delb^2 \Del^2$
acting on the chiral superfield to the invertible operator
\beq
i\Box_+ ~\equiv~ 
\Del^2 \Bar{\Del}^2 ~+~ \Bar{\Del}^2 \Del^2 ~-~ \Bar{\Del}^{\dot\a} 
\Del^2 \Bar{\Del}_{\dot\a} ~=~ \Box ~-~ i W^{\a} \Del_{\a} ~-~
\frac{i}{2} (\Del^{\a} W_{\a})  
\label{boxscalar4d}
\eeq
where
\beq
\Box \equiv \frac12 \Del^{\a \dot\a} \Del_{\a \dot\a} ~~~.
\eeq
We obtain the scalar propagator  (we interchange $z$ and
$z'$)
\beq
\langle \Phi(z) \Phib (z')\rangle = \Bar{\Del}^2 \frac{1}{\Box_+} \Del^2
e^{-V} \d^{(8)} (z-z') ~~~,
\label{4dprop}
\eeq
where we have used the identity $ \Bar{\Del}^2 \Box_+ = \Box_+
\Bar{\Del}^2$.
The previous expression also gives the covariant scalar propagator for
the $(2,2)$ theory.  In the main text we have also used $\Box_- =
\Box - i \bar{W}^\ad \Delb_\ad - \frac{i}{2} (\Delb^\ad \bar{W}_\ad)$.

\subsection{ $(2,0)$ Scalar Superfield}

We compute the two--dimensional propagators by following a procedure
analogous to the one used in the 4D case. From the explicit expression
(\ref{scalaract}) for the scalar action, we write the functional
integral
\beq
\VEV{\Phib(z)}= \int D \Phi D \Phib \exp[ i\int d^4w \Phib (w) e^V
\Delmm \Phi (w) ] \cdot \Phib (z) ~~~.
\eeq
Now making a change of variable $\Phib \rightarrow  \Phib + \d \Phib$
under which the functional integral doesn't change and taking the 
functional derivative $\d / \d \Phib (z')$, we obtain
\beq
i\int d^4u D_+ \d^{(4)} (u-z') e^V \Delmm \VEV{\Phi(u) \Phib (z)} + 
D_+ \d^{(4)} (z-z')  ~=~ 0 ~~~,
\eeq
or
\beq
i D_+ e^V \Delmm \VEV{\Phi(z') \Phib (z)} ~=~ D_+ \d^{(4)} (z-z') 
~=~ - D_+ \d^{(4)} (z'-z) ~~~.
\eeq
We then multiply by $\Delpd e^{-V}$ on the left to obtain
\beq
i \Delpd \Delp \Delmm \VEV{\Phi(z') \Phib (z)} ~=~ - \Delpd e^{-V} D_+  
\d^{(4)} (z'-z) ~=~ - \Delpd \Delp e^{-V} \d^{(4)} (z' - z) ~~~.
\eeq
The derivative operator on the l.h.s. can be suitably extended to an
invertible operator
\beq
i \Box_+ ~\equiv~ \Delpd \Delp \Delmm ~+~ \Delp \Delmm \Delpd ~=~
i [\, \Box + \frac12 (\Delpd \Bar{W}_{\md} + \Delp W_-) - W_- \Delp
\, ] ~~~,
\label{boxscalar}
\eeq
where
\beq
\Box ~\equiv~ \frac12 ( \Delpp \Delmm + \Delmm \Delpp ) ~~~.
\label{boxcov}
\eeq
Therefore, for the scalar propagator we  obtain (we interchange 
$z$ and $z'$)
\beq
\langle \Phi(z) \Phib (z')\rangle ~=~ \frac{1}{\Box_+} \Delpd \Delp
e^{-V} \d^{(4)} (z-z') ~~~.
\label{prop}
\eeq
Since the  identity  $ \Delpd \Box_+ = \Box_+ \Delpd $ holds,
in
the previous equation one can interchange the two operators and write
\beq
\langle \Phi(z) \Phib (z')\rangle ~=~ \Delpd \frac{1}{\Box_+} \Delp
e^{-V} \d^{(4)} (z-z') ~~~.
\label{scalarprop}
\eeq
We also define 
\beq
i\Box_- ~\equiv ~ \Delmm \Delpd \Delp ~+~ 
\Delp \Delmm \Delpd  ~=~ i [\, \Box + \frac12 (\Delp W_- - \Delpd
\Bar{W}_\md ) + \Bar{W}_\md \Del_\pd \,] ~~~.
\label{scalarboxm}
\eeq
This operator satisfies $ \Delp \Box_- = \Box_- \Delp$.

An alternative way to compute the propagator is to use the standard
procedure by completing the square in the functional integral and 
perform the gaussian integration. Let us consider
\beq
{\cal W} = \int [D\Phi_c D\bar{\Phi}_c ] \exp\{ i\int d^2x d^2\theta
\bar{\Phi}_c
\Delmm \Phi_c - \int d^2x d\theta^+ J_c \Phi_c + \int d^2x d\theta^{\pd}
\bar{\Phi}_c  \bar{J}_c \} ~~,
\eeq
where $\Phi_c$ and $\bar{\Phi}_c$ are covariantly chiral and antichiral 
superfields respectively and $J_c$, $\bar{J}_c$ the corresponding sources
(covariantly chiral and antichiral spinors). 
To compute the functional integral we write
\bea
&& 
\int d^2x d \theta^+ J_c \Phi_c ~=~  \int d^2x d^2 \theta J_c \frac{i}{\Box_+} 
\Delp \Delmm \Phi_c ~~~, \nonumber \\
&&
\int d^2x d \theta^{\pd} \bar{\Phi}_c  \bar{J}_c ~=~ \int d^2x d^2 
\theta \bar{\Phi}_c  \Delmm \Delpd\frac{i}{\Box_-} \bar{J}_c ~~~,
\ena
where we have used the operators (\ref{boxscalar},\ref{scalarboxm}).
Then, we have
\beq
\eqalign{ {~~~~~~~}
{\cal W} ~=~ 
\int [D\Phi_c D\bar{\Phi}_c] \exp \Big\{ i\int d^2x d^2\theta &[\, 
(\bar{\Phi}_c 
- J_c \frac{1}{\Box_+} \Del_+ ) \,  \Delmm \, ( \Phi_c + \Delpd
\frac{1}{\Box_-}\bar{J}_c) \cr &~+~ J_c \frac{1}{\Box_+} \Delp \Delmm \Delpd
\frac{1}{\Box_-} \bar{J}_c \, ] ~\Big\}
}\eeq
If we now define
\beq
\Phi_c' ~\equiv~ \Phi_c \,+\,  \Delpd \frac{1}{\Box_-} \bar{J}_c \quad , \quad
\bar{\Phi}_c' ~\equiv~ \bar{\Phi}_c \,-\, J_c \frac{1}{\Box_+} \Del_+ ~~~.
\eeq
the chirality constraints are maintained, $ \Del_{\pd} \Phi_c' =0$ and
$\bar{\Phi}_c' {\stackrel{\longleftarrow}\Del}_+ = 0$, and we can 
perform the gaussian integral obtaining
\beq
{\cal W} ~\sim~ \exp \Big\{ i\int d^2x d^2\theta J_c \frac{1}{\Box_+} \Delp
\Delmm \Delpd
\frac{1}{\Box_-}  \bar{J}_c \Big\} ~~~,
\eeq
We now compute the propagator using the definition (\ref{definition})
\beq
\eqalign{
\langle \Phi_c(z) \bar{\Phi}_c(z') \rangle &\equiv~ -\frac{ \delta} 
{\delta J_c(z)} \frac{\d}{\delta \bar{J}_c(z')} {\cal W} 
\Big|_{J_c=\bar{J}_c=0} \cr
&=~- i\Delpd \frac{1}{\Box_+} \Delp \Delmm \Delpd 
\frac{1}{\Box_-}  \Del_+ \d^{(4)} (z-z') \cr
&=~ -i \Del_\pd \frac{1}{\Box_+} \Del_+ \Del_\mm \Del_\pd \Del_+
\frac{1}{\Box_-}  \d^{(4)} (z-z') \cr
&=~ \Delpd \frac{1}{\Box_+}\Delp  \d^{(4)} (z-z') ~~~.
}
\label{covprop}
\eeq
Going from the first to the second line we get a minus sign from
interchanging $J_c$ with $\frac{\d}{\d \bar{J}_c}$ (remember that 
$J_c$ and $\bar{J}_c$ are spinors) and a minus sign from the functional 
derivative as given in (\ref{der2}), resulting in no change in sign.

Now, remembering that in chiral representation $\Phi_c = \Phi$ and 
$\bar{\Phi}_c = \bar{\Phi} e^V$, from eq. (\ref{covprop}) it is easy to 
infer the result (\ref{scalarprop}).

\subsection{ Spinor Propagator}

We now consider a chiral spinor coupled to Yang--Mills fields, described 
by the action (\ref{spinoract}) and compute the propagator by following
the
same procedure of the scalar case. We start with
\beq
\VEV{\chib_{\md}(z)} ~=~ \int D \chi_- D \chib_{\md} \exp[ \int d^4w
\chib_{\md} (w) e^V \chi_- (w) ] \cdot \chib_{\md} (z) ~~~.
\eeq
and perform a shift $\chib_{\md} \rightarrow \chib_{\md} + \d
\chib_{\md}$.
We obtain
\beq
\eqalign{ {~~~~~~~~}
0~=~ \int D \chi_- D \chib_{\md} &\exp[\,\int d^4w \chib_{\md} (w)
e^V
\chi_- (w) ]\, \times \cr 
&{~~~~~}[\int d^4u \d \chib_{\md} (u) e^V \chi_-(u) \chib_{\md} (z) +
\d
\chib_{\md} (z) \, ] ~~~. }\eeq
Taking the functional derivative $\d/ \d \chib_{\md}(z')$ and following
the same steps as before, we obtain
\beq
D_+ e^V \langle \chi_-(z') \chib_{\md} (z) \rangle ~=~ -D_+ \d^{(4)} (z' -z)
~~~.
\eeq
or (multiplying by $e^{-V}$ and exchanging $z$ and $z'$)
\beq
\Delp \langle \chi_-(z) \chib_{\md} (z') \rangle ~=~ 
-\Delp e^{-V} \d^{(4)} (z -z')
~~~. 
\eeq
Then we apply $\Delpd \Delmm $. Defining the invertible operator 
(note that it is
different from the corresponding one in (C.13)
\beq
i \Box_+ ~\equiv~ \Delpd \Delmm \Delp + \Delmm \Delp \Delpd ~=~ i [\, 
\Box + \frac12 ( \Delp W_- - \Delpd \Bar{W}_{\md} ) - W_- \Delp \,]
~~~,  
\label{spinorbox}
\eeq
where $\Box$ is given in (\ref{boxcov}), we  obtain
\beq
\langle \chi_-(z) \bar{\chi}_{\md}(z') \rangle ~=~ \Delpd
\frac{i}{\Box_+} \Delmm \Delp e^{-V} \d^{(4)} (z -z') ~~~.
\label{spinorprop}
\eeq
In the text we have also made use of the following operator
\beq
i \Box_- ~\equiv~ \Del_\pd \Del_\mm \Del_+ ~+~ \Del_+ \Del_\pd \Del_\mm
~=~ i\, [\, \Box - \frac12 ( \Del_\pd \Bar{W}_\md + \Del_+ W_- )
+ \Bar{W}_\md \Del_\pd \,]  ~~~. 
\label{spinorboxm}
\eeq

\sect{Evaluation of the 4D consistent anomaly terms}

We give here some details in the derivation of  (\ref{xterm}) 
from the last line in (\ref{expression})
\beq
-  \int_0^1 dy \int d^8z  \int_0^1 d \b \frac{1}{M^2} g^{-1}h_2
 e^{\b \Box_+/M^2} \Bar{\Del}^2 h_1 e^{(1-\b )\Box_-/M^2}
\Del^2 \d^{(8)} (z-z') ~~~.
\eeq
with  $h_2 = g^{-1}\d g$ and $h_2 = g^{-1}\partial_y g$.
As described in the main text, we  write 
\beq
\d^{(8)}(z-z') = \frac{M^4}{(2\pi)^4}
\int d^4k e^{iMk(x-x') } \d^{(4)}(\theta - \theta')
\eeq
and we pull the exponential
through the various derivatives, after which we take the limit $x' \to x$. 
We obtain
(aside from some explicit $\theta$  terms that can be argued away \cite{OHS}),
\bea
&&- \frac{M^2}{(2\pi)^4}\int_0^1 dy \int d^8z  \int d^4k \int_0^1 d \beta  ~ 
h_2e^{-\b [k^2 +\cdots +i{\cal W}^\a \Del_\a/M^2 
+\frac{i}{2} (\Del^\a {\cal W}_\a)/M^2]} \nonumber\\
&&~~~~~~~~~~~~~ \cdot\Delb^2 h_1 e^{-(1-\b) [k^2 +\cdots +
i\Tilde{\cal W}^\ad \Del_\ad /M^2 +\frac{i}{2}  (\Del^\ad 
\Tilde{\cal W}_\ad)/M^2]} \Del^2 \d^{(4)}(\theta - \theta')\\
&&\Longrightarrow
\frac{i}{(2\pi)^4} \int d^4k e^{-k^2} \, \int_0^1 dy \int d^8z    
\int_0^1 d \beta \left\{ \b [h_2 {\cal W}^\a \Del_\a+
\frac{i}{2} (\Del^\a {\cal W}_\a) ] \Delb^2 h_1 \Del^2\right. \nonumber \\
&&\left.~~~~~~~~~~~~~~~~~~~~~~~~~~~
+(1-\b) h_2 \Delb^2h_1[\Tilde{\cal W}^\ad \Del_\ad +
\frac{i}{2} (\Del^\ad \Tilde{\cal W}_\ad)] \Del^2 \right\}  
\d^{(4)}(\theta - \theta')
\nonumber 
\eea
where we have expanded the exponential, discarded a divergent term which
is cancelled by the $z\leftrightarrow z''$ term in the original expression for
the anomaly, and kept terms that can contribute in the limit 
$\theta \to \theta'$.
The $k$ and $\b$ integrations can now be performed.

After using two factors of $\Del_\a$ and $\Delb_\ad$ to remove the $\theta$'s,
and subtracting the contribution with $z$ and $z''$ interchanged (which is
equivalent to interchanging $h_1$ and $h_2$), we obtain
\bea
\frac{i}{8\pi^2} \,  
\int_0^1 dy \int d^8z \, \Big[ h_2 {\cal W}^\a (\Del_\a h_1) 
+\frac{1}{2}h_2
(\Del^\a {\cal W}_\a)h_1 -h_2 \Delb^\ad(h_1 \Tilde{\cal W}_\ad) +\frac{1}{2}
h_2h_1 (\Delb^\ad \Tilde{\cal W}_\ad)  \nonumber\\
- h_1 {\cal W}^\a (\Del_\a h_2) - \frac{1}{2}h_1
(\Del^\a {\cal W}_\a)h_2 + h_1 \Delb^\ad(h_1 \Tilde{\cal W}_\ad) - \frac{1}{2}
h_1h_2 (\Delb^\ad \Tilde{\cal W}_\ad) \Big]  \nonumber\\
\eea
Using the cyclicity of the (group theory) trace, this expression can be
simplified to the one given in (\ref{xterm}).

\end{document}